\documentclass[preprint,showpacs,preprintnumbers,superscriptaddress,nofootinbib]{revtex4}

\usepackage{graphicx}
\usepackage{dcolumn}
\usepackage{bm}
\usepackage{amsmath}
\usepackage{amssymb}
\usepackage{url}
\usepackage[normalem]{ulem}
\usepackage{xcolor}
\usepackage{listings} 
\newcommand{\beq}{\begin{equation}}
\newcommand{\eeq}{\end{equation}}
\newcommand{\bea}{\begin{eqnarray}}
\newcommand{\eea}{\end{eqnarray}}
\newcommand{\real}{{\sf I}\kern-.12em{\sf R}}
\newcommand{\comp}{{\sf I}\kern-.50em{\sf C}}
\newcommand{\unity}{{\sf I}\kern-.54em{\sf 1}}

\newcommand{\stringa}{\ttfamily\lstinline}
\def\cod#1{{\stringa!#1!}}

\def\spose#1{\hbox to 0pt{#1\hss}}
\def\ltapprox{\mathrel{\spose{\lower 3pt\hbox{$\mathchar"218$}}
 \raise 2.0pt\hbox{$\mathchar"13C$}}}

\begin{document}

\title{Dihadron production at the LHC:
       full next-to-leading BFKL calculation}
\author{Francesco G. Celiberto}
\affiliation{Dipartimento di Fisica dell'Universit\`a della Calabria \\
I-87036 Arcavacata di Rende, Cosenza, Italy}
\affiliation{INFN - Gruppo collegato di Cosenza, I-87036 Arcavacata di Rende,
Cosenza, Italy}
\author{Dmitry Yu. Ivanov}
\affiliation{Sobolev Institute of Mathematics, 630090 Novosibirsk, Russia}
\affiliation{Novosibirsk State University, 630090 Novosibirsk, Russia}
\author{Beatrice Murdaca}
\affiliation{INFN - Gruppo collegato di Cosenza, I-87036 Arcavacata di Rende,
Cosenza, Italy}
\author{Alessandro Papa}
\affiliation{Dipartimento di Fisica dell'Universit\`a della Calabria \\
I-87036 Arcavacata di Rende, Cosenza, Italy}
\affiliation{INFN - Gruppo collegato di Cosenza, I-87036 Arcavacata di Rende,
Cosenza, Italy}

\date{\today}

\begin{abstract}
The study of the inclusive production of a pair of charged light hadrons 
(a ``dihadron'' system) featuring high transverse momenta 
and well separated in rapidity represents a clear channel for the test
of the BFKL dynamics at the Large Hadron Collider (LHC). 
This process has much in common with the well known Mueller-Navelet jet
production; however, hadrons can be detected at much smaller values of the
transverse momentum than jets, thus allowing to explore an additional
kinematic range, supplementary to the one studied with Mueller-Navelet jets. 
Furthermore, it makes it possible to constrain not only the parton densities 
(PDFs) for the initial proton, but also the parton fragmentation functions
(FFs) describing the detected hadron in the final state.
Here, we present the first full NLA BFKL analysis for cross sections and
azimuthal angle correlations for dihadrons produced in the LHC kinematic
ranges. 
We make use of the Brodsky-Lapage-Mackenzie (BLM) optimization method to set
the values of the renormalization scale and study the effect 
of choosing different values for the factorization scale. 
We also gauge the uncertainty coming from the use of different PDF and
FF parametrizations. 
\end{abstract}

\maketitle

\section{Introduction}
\label{introd}

Semi-hard processes in the large center-of-mass energy limit represent
a unique arena to test strong interactions in kinematic regimes so far
unexplored, the high luminosity and the record energies 
of hadronic processes at the Large Hadron Collider (LHC) 
providing with a wealth of useful data. 
In the kinematical regime $s\gg |t|$, known as Regge limit,  
fixed-order calculations in perturbative QCD based on collinear factorization
miss the effect of large energy logarithms, entering the perturbative series
with a power increasing with the order and thus compensating the smallness
of the coupling $\alpha_s$. The Balitsky--Fadin--Kuraev--Lipatov (BFKL)
approach~\cite{BFKL} serves as the most powerful tool 
to perform the all-order resummation of these large energy logarithms 
both in the leading approximation (LLA), 
which means all terms proportional to $(\alpha_s\ln(s))^n$,
and the next-to-leading approximation (NLA),
which means all terms proportional to $\alpha_s(\alpha_s\ln(s))^n$.
In the BFKL formalism, it is possible to express the cross section of
an LHC process falling in the domain of perturbative QCD 
as the convolution between two impact factors, which describe 
the transition from each colliding proton to the respective 
final state object, and a process-independent Green's function.
The BFKL Green's function obeys an integral equation, whose
kernel is known at the next-to-leading order (NLO) both for forward
scattering ({\it i.e.} for $t=0$ and color singlet in the
$t$-channel)~\cite{Fadin:1998py,Ciafaloni:1998gs} 
and for any fixed (not growing with energy)
momentum transfer $t$ and any possible two-gluon color state in the
$t$-channel~\cite{Fadin:1998jv,FG00,FF05}.

The too low $\sqrt{s}$, together with small rapidity intervals
among the tagged objects in the final state, 
had been so far the weakness point of the search for BFKL effects. 
Furthermore, too inclusive observables were considered. 
A striking example is the growth of the hadron structure functions 
at small Bjorken-$x$ values in Deep Inelastic Scattering (DIS).
Although NLA BFKL predictions for the structure function $F_{2,L}$ 
have shown a good agreement with the HERA 
data~\cite{Hentschinski:2012kr,Hentschinski:2013id}, 
other approaches can fit these data. 
The LHC record energy, together with the good resolution 
in azimuthal angles of the particle detectors, can address these issues: 
on one side larger rapidity intervals in the final state are reachable, 
allowing us to study a kinematic regime where 
it is possible to disentangle the BFKL dynamics from other resummations;
on the other side there is enough statistics 
to define and investigate more exclusive observables, 
which can, in principle, be only described by the BFKL framework.

With this aim, the production of two jets featuring transverse momenta much
larger than $\Lambda^2_{\rm QCD}$ and well separated in rapidity, 
known as Mueller-Navelet jets, was proposed~\cite{Mueller:1986ey} 
as a tool to investigate semi-hard parton scatterings at a hadron collider. 
This reaction represents a unique venue where two main resummations, collinear
and BFKL ones, play their role at the same time in the context of perturbative
QCD. On one hand, the rapidity ranges in the final state are large enough 
to let the NLA BFKL resummation of the energy logarithms come into play. 
The process-dependent part of the information needed to build up 
the cross section is encoded in the impact factors
(the so-called ``jet vertices''), which are known 
up to NLO~\cite{Bartels:2001ge,Bartels:2002yj,Caporale:2011cc,Ivanov:2012ms,Colferai:2015zfa}.
On the other hand, the jet vertex can be expressed, within collinear
factorization at the leading twist, as the convolution of the parton
distribution function (PDF) of the colliding proton, obeying the standard
DGLAP evolution~\cite{DGLAP}, with the hard process describing the transition
from the parton emitted by the proton to the forward jet in the final state.

A large number of numerical analyses~\cite{Colferai:2010wu,Angioni:2011wj,Caporale:2012ih,Ducloue:2013wmi,Ducloue:2013bva,Caporale:2013uva,Ducloue:2014koa,Caporale:2014gpa,Ducloue:2015jba,Caporale:2015uva,Celiberto:2015yba,Celiberto:2016ygs,Chachamis:2015crx} has appeared so far, which have been devoted to NLA BFKL
predictions for the Mueller-Navelet jet production process. 
All these studies are involved in calculating cross sections 
and azimuthal angle correlations~\cite{DelDuca:1993mn,Stirling:1994he} 
between the two measured jets, {\it i.e.} average values of $\cos{(n \phi)}$, 
where $n$ is an integer and $\phi$ is the angle in the azimuthal plane
between the direction of one jet and the direction opposite to the other jet,
and ratios of two such cosines~\cite{Vera:2006un,Vera:2007kn}. 
Recently~\cite{Khachatryan:2016udy}, 
the CMS Collaboration presented the first measurements of the azimuthal 
correlation of the Mueller-Navelet jets at $\sqrt{s}=7$~{TeV} at LHC.  
Further experimental studies of the Mueller-Navelet jets 
at higher LHC energies and larger rapidity intervals, including also the
effects of using asymmetrical cuts for the jet transverse momenta, are
expected.  

In order to uncover the dynamical mechanisms behind partonic interactions
in the Regge limit, new observables, sensitive to the BFKL dynamics and
less inclusive than the Mueller-Navelet ones, need to be proposed and
considered in the next LHC analyses. 
An interesting option, the detection of three jets, well separated in rapidity
from each other, has been proposed in
Refs.~\cite{Caporale:2015vya,Caporale:2016soq} and recently investigated with
NLA BFKL accuracy in Ref.~\cite{Caporale:2016zkc}.
Its natural extension, the four-jet production process, has been proposed in
Ref.~\cite{Caporale:2015int} and studied in Ref.~\cite{Caporale:2016xku}.

\begin{figure}[t]
 \centering
 \includegraphics[scale=0.5]{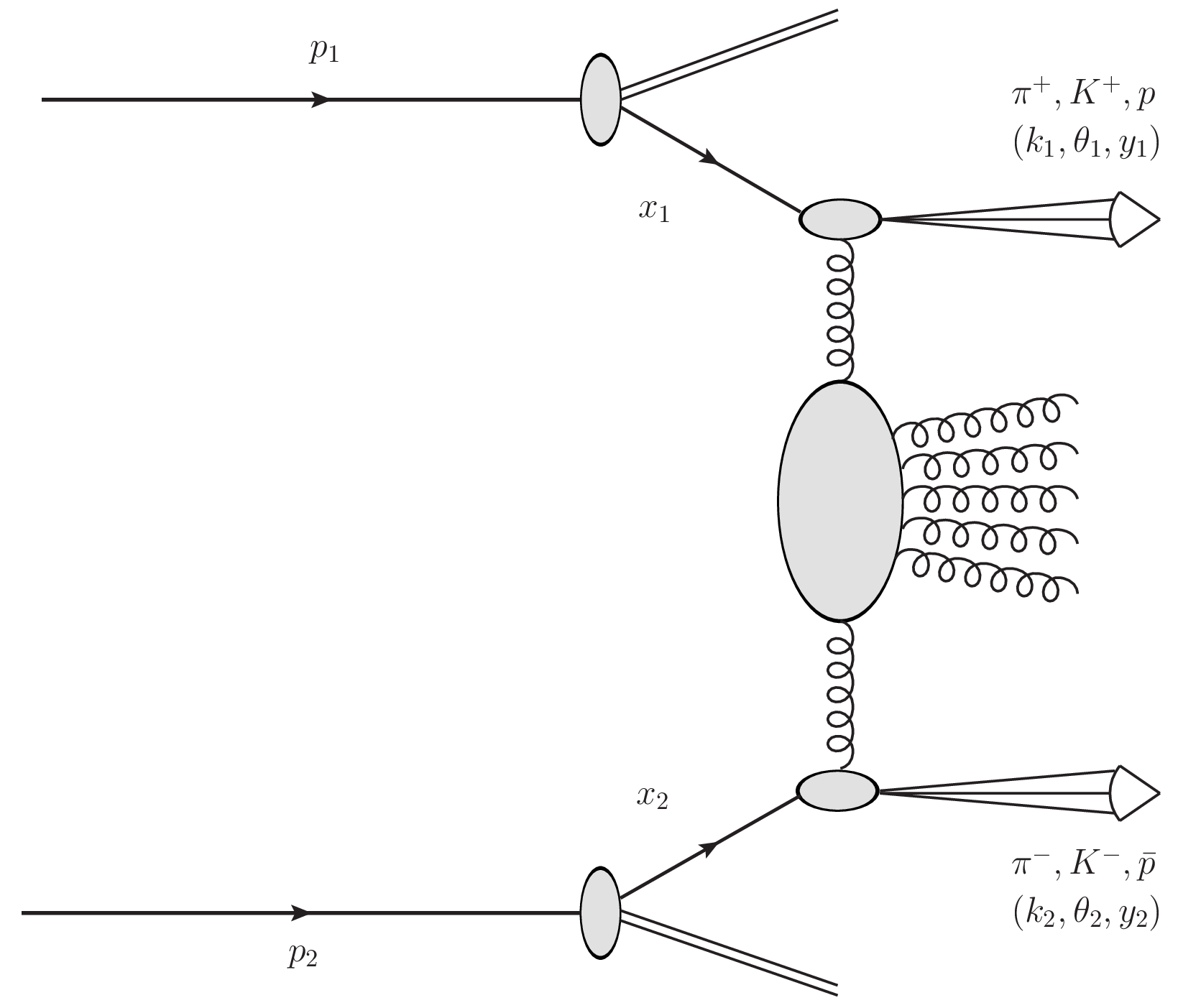}
 \caption[]
 {Inclusive dihadron production process in multi-Regge kinematics.}
 \label{fig:di-hadron}
 \end{figure}

In a recent paper~\cite{Celiberto:2016hae} we suggested a novel possibility,
{\it i.e.} the inclusive dihadron production 
\begin{eqnarray}
\label{process}
{\rm p}(p_1) + {\rm p}(p_2) 
\to 
{\rm h}_1(k_1) + {\rm h}_2(k_2)+ {\rm X} \;,
\end{eqnarray}
when the two charged light hadrons: $\pi^{\pm}, K^{\pm}, p,\bar p$ 
with high transverse momenta and separated by a large interval of rapidity,
together with an undetected hadronic system X, are produced in the final state 
(see Fig.~\ref{fig:di-hadron} for a schematic view).

This process is similar to the Mueller-Navelet jet production and shares
with it the underlying theoretical framework, the only obvious difference
lying in the vertices describing the dynamics in the proton fragmentation
region: instead of the proton-to-jet vertex, the vertex for the
proton to identified hadron transition is needed.
Such a vertex was considered in~\cite{hadrons} within NLA: it was shown
there that ultraviolet divergences are taken care of by the renormalization
of the QCD coupling, soft and virtual infrared divergences cancel each
other, whereas the surviving infrared collinear ones are compensated by
the collinear counterterms related with the renormalization of PDFs for
the initial proton and parton fragmentation functions (FFs) describing the
detected hadron in the final state within collinear factorization.
\footnote{The identified hadron production vertex  in the NLA  was found
  within the shockwave approach (or Color Glass Condensate effective theory)
  in~\cite{Chirilli:2012jd}. It  was  used there  to  study the
  single inclusive particle production at forward rapidities in proton-nucleus
  collisions; for recent developments of this line of research, see
  also~\cite{Iancu:2016vyg}. Unfortunately, the comparison  between  the
  results of~\cite{Chirilli:2012jd} and those of~\cite{hadrons} is not
  simple and straightforward, since the distribution of radiative corrections
  between the kernel and the impact factor is different in the shockwave and the
  BFKL frameworks. Nontrivial kernel and impact factor transformations are
  required for such a comparison. It certainly deserves a separate study,
  and the consideration of the process~(\ref{process}) within both the
  shockwave and the BFKL resummation schemes seems the best possibility
   to this purpose. }
Hence, infrared-safe NLA predictions for observables related with this
process are amenable, thus making this process an additional 
clear channel to test the BFKL dynamics at the LHC. The reaction~(\ref{process})
can be considered complementary to Mueller-Navelet jet production, since
hadrons can be detected at the LHC at much smaller values of the transverse
momentum than jets, thus giving access to a kinematic range outside the
reach of the Mueller-Navelet channel.

Note that the inclusive dihadron production was analysed by
CMS~\cite{Khachatryan:2010gv,Khachatryan:2015lva} and ATLAS~\cite{Aad:2015gqa}
Collaborations at different LHC energies. The focus was put on two-particle
azimuthal angle and rapidity correlations for charged hadrons at low and
medium transverse momenta. Here we suggest to analyze this reaction in the
region of larger transverse momenta, where data could be confronted with
perturbative QCD predictions.

In Ref.~\cite{Celiberto:2016hae} we gave the first predictions for cross
sections and azimuthal angle correlations of the process~(\ref{process}) 
in an approximated way, since we neglected, for the sake of simplicity, the
NLA corrections of the hadron vertices. It is known that the inclusion of
NLA terms has a large impact on the theory predictions for the
Mueller-Navelet jet cross sections and the jet 
azimuthal angle distributions. Similar features are expected 
also for our case of inclusive dihadron production.  As for  Mueller-Navelet
jets, the inclusion of full NLA effects in the process~(\ref{process}) is very
important in order to have a control on the accuracy of predictions, in
particular on effects related with the choice of  the
renormalization scale $\mu_R$ and the factorization scale $\mu_F$.  

The main aim of this paper is to extend and complete the analysis
done in~Ref.~\cite{Celiberto:2016hae} by giving full NLA predictions 
at $\sqrt{s} = 7,13$ TeV and considering two distinct ranges 
for the rapidity interval $Y$ between the two hadrons:
$Y \leq 4.8$ and $Y \leq 9.4$.
It is well known that, even after the account of the NLA effects, predictions
within BFKL resummation still suffer  from large ambiguities in the choice
of scales. As an idea for the renormalization scale choice setting  
we adopt the Brodsky-Lepage-Mackenzie (BLM) scheme~\cite{Brodsky:1982gc}. 
In BLM the renormalization scale ambiguity is eliminated by 
absorbing the non-conformal, proportional to the QCD $\beta_0$-function,
terms into the running coupling. Such approach was successfully used, 
first in~\cite{Ducloue:2013bva}, for a satisfactory description of the LHC
data on the azimuthal correlations of Mueller-Navelet
jets~\cite{Khachatryan:2016udy}, obtained by the CMS collaboration.

As for the factorization scale, we chose either to fix it equal to 
the renormalization scale, $\mu_F = \mu_R = \mu_R^{\rm BLM}$, or we use a scheme
with two separate values of the factorization scale and fix them at the
transverse momentum of one or the other of the two detected hadrons, 
$(\mu_F)_{1,2}= |\vec k_{1,2}|$, depending on which of the two
vertices is considered.

The summary of the paper is as follows: in Section~1 we present the theoretical
framework and sketch the derivation of our predictions; in Section~2 we
show and discuss the results of our numerical analysis; finally, in Section~3,
we draw our conclusions and give some outlook.

\section{Theoretical framework}
\label{theory}

The process under investigation (see~(\ref{process}) and
Fig.~\ref{fig:di-hadron}) is the inclusive production of a pair of identified
hadrons featuring large transverse momenta, 
$\vec k_1^2\sim \vec k_2^2 \gg \Lambda^2_{\rm QCD}$ 
and separated by a large rapidity interval in high-energy proton-proton
collisions. The protons' momenta $p_1$ and $p_2$ are taken as Sudakov
vectors satisfying $p^2_1= p^2_2=0$ and $2 (p_1p_2) = s$, 
so that the momentum of each hadron can be decomposed as
\bea
k_1&=& \alpha_1 p_1+ \frac{\vec k_1^2}{\alpha_1 s}p_2+k_{1\perp} \ , \quad
k_{1\perp}^2=-\vec k_1^2 \ , \nonumber \\
k_2&=& \alpha_2 p_2+ \frac{\vec k_2^2}{\alpha_2 s}p_1+k_{2\perp} \ , \quad
k_{2\perp}^2=-\vec k_2^2 \ .
\label{sudakov}
\eea

In the center of mass system, the hadrons' longitudinal momentum fractions 
$\alpha_{1,2}$ are connected to the respective 
rapidities through the relations
$y_1=\frac{1}{2}\ln\frac{\alpha_1^2 s}
{\vec k_1^2}$, and $y_2=\frac{1}{2}\ln\frac{\vec k_2^2}{\alpha_2^2 s}$, 
so that $dy_1=\frac{d\alpha_1}{\alpha_1}$, $dy_2=-\frac{d\alpha_2}{\alpha_2}$,
and $Y=y_1-y_2=\ln\frac{\alpha_1\alpha_2 s}{|\vec k_1||\vec k_2|}$, here the
space part of the four-vector $p_{1\parallel}$ being taken positive.

In QCD collinear factorization the cross section 
of the process~(\ref{process}) reads
\beq
\frac{d\sigma}{d\alpha_1d\alpha_2d^2k_1d^2k_2}
=\sum_{a,b=q,{\bar q},g}\int_0^1 dx_1 \int_0^1 dx_2\ f_a\left(x_1,\mu_F\right)
\ f_b\left(x_2,\mu_F\right)
\frac{d{\hat\sigma}_{a,b}\left(\hat s,\mu_F\right)}
{d\alpha_1d\alpha_2d^2k_1d^2k_2}\;,
\eeq
where the $a, b$ indices specify the parton types 
(quarks $q = u, d, s, c, b$;
antiquarks $\bar q = \bar u, \bar d, \bar s, \bar c, \bar b$; 
or gluon $g$),
$f_a\left(x, \mu_F \right)$ denotes the initial proton PDFs; 
$x_{1,2}$ are the longitudinal fractions of the partons 
involved in the hard subprocess, 
while $\mu_F$ is the factorization scale; 
$d\hat\sigma_{a,b}\left(\hat s \right)$ is
the partonic cross section and
$\hat s \equiv x_1x_2s$ is the squared center-of-mass energy of the
parton-parton collision subprocess.

In the BFKL approach the cross section can be presented 
(see Ref.~\cite{Caporale:2012ih} for the details of the derivation)
as the Fourier sum of the azimuthal coefficients ${\cal C}_n$, 
having so:
\beq
\frac{d\sigma}
{dy_1dy_2\, d|\vec k_1| \, d|\vec k_2|d\phi_1 d\phi_2}
=\frac{1}{(2\pi)^2}\left[{\cal C}_0+\sum_{n=1}^\infty  2\cos (n\phi )\,
{\cal C}_n\right]\, ,
\eeq
where $\phi=\phi_1-\phi_2-\pi$, with $\phi_{1,2}$ are the two hadrons' 
azimuthal angles, while $y_{1,2}$ and $\vec k_{1,2}$ are their
rapidities and transverse momenta, respectively. 
The $\phi$-averaged cross section ${\cal C}_0$ 
and the other coefficients ${\cal C}_{n\neq 0}$ are given by
\beq\nonumber
{\cal C}_n \equiv \int_0^{2\pi}d\phi_1\int_0^{2\pi}d\phi_2\,
\cos[n(\phi_1-\phi_2-\pi)] \,
\frac{d\sigma}{dy_1dy_2\, d|\vec k_1| \, d|\vec k_2|d\phi_1 d\phi_2}\;
\eeq
\beq\nonumber
= \frac{e^Y}{s}
\int_{-\infty}^{+\infty} d\nu \, \left(\frac{\alpha_1 \alpha_2 s}{s_0}
\right)^{\bar \alpha_s(\mu_R)\left[\chi(n,\nu)+\bar\alpha_s(\mu_R)
\left( \bar\chi(n,\nu)+\frac{\beta_0}{8 N_c}\chi(n,\nu)\left(-\chi(n,\nu)
+\frac{10}{3}+\ln\frac{\mu_R^4}{\vec k_1^2\vec k_2^2}\right)\right)\right]}
\eeq
\beq\nonumber
\times \alpha_s^2(\mu_R) c_1(n,\nu,|\vec k_1|, \alpha_1)
c_2(n,\nu,|\vec k_2|,\alpha_2)\,
\eeq
\beq\label{Cm}
\times \left[1
+\alpha_s(\mu_R)\left(\frac{c_1^{(1)}(n,\nu,|\vec k_1|,
\alpha_1)}{c_1(n,\nu,|\vec k_1|, \alpha_1)}
+\frac{c_2^{(1)}(n,\nu,|\vec k_2|, \alpha_2)}{c_2(n,\nu,|\vec k_2|,
\alpha_2)}\right)\right.
\eeq
\beq\nonumber
\left. + \bar\alpha_s^2(\mu_R) \ln\frac{\alpha_1 \alpha_2 s}{s_0}
\frac{\beta_0}{8 N_c}\chi(n,\nu) \left(2\ln \vec k_1^2 \vec k_2^2
+i\frac{d\ln\frac{c_1(n,\nu)}{c_2(n,\nu)}}{d\nu}\right)\right]\;.
\eeq
\\
Here $\bar \alpha_s(\mu_R) \equiv \alpha_s(\mu_R) N_c/\pi$, with
$N_c$ the number of colors
\beq
\beta_0=\frac{11}{3} N_c - \frac{2}{3}n_f
\eeq
is the first coefficient of the QCD $\beta$-function, where $n_f$ is the number
of active flavors.
\beq
\chi\left(n,\nu\right)=2\psi\left(1\right)-\psi\left(\frac{n}{2}
+\frac{1}{2}+i\nu \right)-\psi\left(\frac{n}{2}+\frac{1}{2}-i\nu \right)
\eeq
is the leading-order (LO) BFKL characteristic function,
$c_{1,2}(n,\nu)$ are the LO impact factors in the $\nu$-repre\-sen\-ta\-tion, 
that are given as an integral in the parton fraction $x$, containing
the PDFs of the gluon and of the different quark/antiquark
flavors in the proton, and the FFs of the detected hadron,
\bea
c_1(n,\nu,|\vec k_1|,\alpha_1) &=& 2 \sqrt{\frac{C_F}{C_A}}
(\vec k_1^2)^{i\nu-1/2}\,\int_{\alpha_1}^1\frac{dx}{x}
\left( \frac{x}{\alpha_1}\right)
^{2 i\nu-1} 
\nonumber \\
\label{c1}
&\times&\left[\frac{C_A}{C_F}f_g(x)D_g^h\left(\frac{\alpha_1}{x}\right)
+\sum_{a=q,\bar q}f_a(x)D_a^h\left(\frac{\alpha_1}{x}\right)\right]
\eea
and
\beq
\label{c2}
c_2(n,\nu,|\vec k_2|,\alpha_2)=
\biggl[c_1(n,\nu,|\vec k_2|,\alpha_2)\biggr]^* \;,
\eeq
while
\bea\label{c11}
c_1^{(1)}(n,\nu,|\vec k_1|,\alpha_1)&=&
2\sqrt{\frac{C_F}{C_A}}
\left(\vec k_1^2\right)^{i\nu-\frac{1}{2}}\frac{1}{2\pi}
\int_{\alpha_1}^1\frac{d x}{x}
\int_{\frac{\alpha_1}{x}}^1\frac{d\zeta}{\zeta}
\left(\frac{x\zeta}{\alpha_1}\right)^{2i\nu-1}
\\
\nonumber
&\times&
\left[
\frac{C_A}{C_F}f_g(x)D_g^h\left(\frac{\alpha_1}{x\zeta}\right)C_{gg}
\left(x,\zeta\right)+\sum_{a=q,\bar q}f_a(x)D_a^h
\left(\frac{\alpha_1}{x\zeta}
\right)C_{qq}\left(x,\zeta\right)
\right.
\\
\nonumber
&+&\left.D_g^h\left(\frac{\alpha_1}{x\zeta}\right)
\sum_{a=q,\bar q}f_a(x)C_{qg}
\left(x,\zeta\right)+\frac{C_A}{C_F}f_g(x)\sum_{a=q,\bar q}D_a^h
\left(\frac{\alpha_1}{x\zeta}\right)C_{gq}\left(x,\zeta\right)
\right]\, ,
\eea
and 
\beq\label{c21}
c_2^{(1)}(n,\nu,|\vec k_2|,\alpha_2)=\biggl[c_1^{(1)}(n,\nu,|\vec k_2|,
\alpha_2)\biggr]^*
\eeq
are the NLO impact factor corrections in the $\nu$-representation. The
expressions for them can be derived from the last two lines 
of Eq.~(4.58) in Ref.~\cite{hadrons}.
It is known~\cite{Caporale:2015uva} that contributions to the NLO impact
factors  that are proportional to the QCD $\beta_0$-function are
universally expressed in terms of the LO impact factors of the considered
process, through the function $f\left(\nu\right)$, defined as follows:
\begin{equation}\nonumber
\label{nu}
2\ln \mu_R^2+i\frac{d\ln\frac{c_1(n,\nu)}{c_2(n,\nu)}}{d\nu}
=\ln\frac{\mu_R^4}{\vec k_1^2\vec k_2^2}
\end{equation}
\begin{equation}\nonumber
-2 \frac{\int_{\alpha_1}^1\frac{d x}{x}\left(\frac{x}{\alpha_1}
\right)^{2 i \nu-1}\log\left(\frac{x}{\alpha_1}\right)\left[\frac{C_A}{C_F}
f_g(x)D_g^h\left(\frac{\alpha_1}{x}\right)+\sum_{a=q,\bar q}f_a(x)D_a^h
\left(\frac{\alpha_1}{x}\right)\right]}
{\int_{\alpha_1}^1\frac{d x}{x}\left( \frac{x}{\alpha_1}\right)^{2 i \nu-1}
\left[\frac{C_A}{C_F}f_g(x)D_g^h\left(\frac{\alpha_1}{x}\right)+\sum_{a=q,\bar q}
f_a(x)D_a^h\left(\frac{\alpha_1}{x}\right)\right]}
\end{equation}
\begin{equation}\nonumber
-2  \frac{\int_{\alpha_2}^1\frac{d x}{x}\left( \frac{x}{\alpha_2}
\right)^{-2 i \nu-1}\log\left(\frac{x}{\alpha_2}\right)\left[\frac{C_A}{C_F}
f_g(x)D_g^h\left(\frac{\alpha_2}{x}\right)+\sum_{a=q,\bar q}f_a(x)D_a^h
\left(\frac{\alpha_2}{x}\right)\right]}
{\int_{\alpha_2}^1\frac{d x}{x}\left( \frac{x}{\alpha_2}\right)^{-2 i \nu-1}
\left[\frac{C_A}{C_F}f_g(x)D_g^h\left(\frac{\alpha_2}{x}\right)+\sum_{a=q,\bar q}
f_a(x)D_a^h\left(\frac{\alpha_2}{x}\right)\right]}
\end{equation}
\begin{equation}
\equiv \ln\frac{\mu_R^4}{\vec k_1^2\vec k_2^2} + 2f(\nu)\, .
\end{equation}

It is known~\cite{Brodsky:1996sgBrodsky:1997sdBrodsky:1998knBrodsky:2002ka}, 
that in the BLM approach applied to semihard processes, 
we need to perform a finite renormalization 
from the $\overline{\rm MS}$ to the physical MOM scheme, whose definition
is related to the 3-gluon vertex being a key ingredient of the BFKL
resummation. So, we have
\beq{}
\alpha_s^{\overline{\rm MS}}=\alpha_s^{\rm MOM}\left(1+\frac{\alpha_s^{\rm MOM}}{\pi}T
\right)\;,
\eeq
with $T=T^{\beta}+T^{\rm conf}$,
\beq{}
T^{\beta}=-\frac{\beta_0}{2}\left( 1+\frac{2}{3}I \right)\, ,
\eeq
\[ 
T^{\rm conf}= \frac{3}{8}\left[ \frac{17}{2}I +\frac{3}{2}\left(I-1\right)\xi
+\left( 1-\frac{1}{3}I\right)\xi^2-\frac{1}{6}\xi^3 \right] \;,
\]
where $I=-2\int_0^1dx\frac{\ln\left(x\right)}{x^2-x+1}\simeq2.3439$ and $\xi$
is the gauge parameter of the MOM scheme, fixed at zero in the following.
The optimal scale $\mu_R^{\rm BLM}$ is the value of $\mu_R$ that makes 
the $\beta_0$-dependent part in the expression for the observable
of interest vanish.  
In~\cite{Caporale:2015uva} some of us showed that terms proportional 
to the QCD $\beta_0$-function are present not only in the NLA BFKL kernel, 
but also in the expressions for the NLA impact factor. This leads 
to a non-universality of the BLM scale and to its dependence on the 
energy of the process. 

Finally, the condition for the BLM scale setting was found to be 
\[
C^{\beta}_n
\propto \!\!
\int_{y_{1,\rm min}}^{y_{1,\rm max}}dy_1
\int_{y_{2,\rm min}}^{y_{2,\rm max}}dy_2\int_{k_{1,\rm min}}^{\infty}dk_1
\int_{k_{2,\rm min}}^{\infty}dk_2
\!\! 
\int\limits^{\infty}_{-\infty} \!\!d\nu\,e^{Y \bar \alpha^{\rm MOM}_s(\mu^{\rm BLM}_R)\chi(n,\nu)}
c_1(n,\nu)c_2(n,\nu)
\]
\[
\left[\frac{5}{3}
+\ln \frac{(\mu^{\rm BLM}_R)^2}{|\vec k_1|
|\vec k_2|} +f(\nu)-2\left( 1+\frac{2}{3}I \right)
\right.
\]
\beq{}
\label{beta0}
\left.
+\bar \alpha^{\rm MOM}_s(\mu^{\rm BLM}_R) Y \: \frac{\chi(n,\nu)}{2}
\left(-\frac{\chi(n,\nu)}{2}+\frac{5}{3}+\ln \frac{(\mu^{\rm BLM}_R)^2}{|\vec k_1|
|\vec k_2|}
+f(\nu)-2\left( 1+\frac{2}{3}I \right)\right)\right]=0 \, .
\eeq
The first term in the r.h.s. of~(\ref{beta0}) originates from the NLA
corrections to the hadron vertices and the second one
(proportional to $\alpha^{\rm MOM}_s$) from the NLA part of the kernel.

\section{Results and Discussion}
\label{results}

\subsection{Integration over the final state phase space}

In order to match the actual LHC kinematical cuts, we integrate the
coefficients over the phase space for two final state hadrons,  
\beq
\label{Cm_int}
C_n= 
\int_{y_{1,\rm min}}^{y_{1,\rm max}}dy_1
\int_{y_{2,\rm min}}^{y_{2,\rm max}}dy_2\int_{k_{1,\rm min}}^{\infty}dk_1
\int_{k_{2,\rm min}}^{\infty}dk_2
\, {\cal C}_n \left(y_1,y_2,k_1,k_2 \right)\, .
\eeq
For the integrations over rapidities we consider two distinct ranges:
\begin{enumerate}
\item 
$y_{1,\rm min}=-y_{2,\rm max}=-2.4$, 
$y_{1,\rm max}=-y_{2,\rm min}=2.4$, 
and $Y \leq 4.8$, \\
typical for the identified hadron detection at LHC; \,
\item 
$y_{1,\rm min}=-y_{2,\rm max}=-4.7$, 
$y_{1,\rm max}=-y_{2,\rm min}=4.7$,
and $Y \leq 9.4$, \\
similar to those used in the CMS Mueller-Navelet jets analysis.
\end{enumerate}
As minimum transverse momenta we choose $k_{1,\rm min}=k_{2,\rm min}=5$~GeV,
which are also realistic values for the LHC. We observe that the minimum
transverse momentum in the CMS analysis~\cite{Khachatryan:2016udy} 
of Mueller-Navelet jet production is much larger, 
$k^{(\rm jet)}_{\rm min}=35$~GeV. In our calculations we use 
the PDF set MSTW 2008 NLO~\cite{Martin:2009iq} 
with two different NLO parameterizations for hadron FFs:  
AKK~\cite{Albino:2008fy} and HKNS~\cite{Hirai:2007cx} 
(see Section~\ref{numerics} for a related discussion). 
In the results presented below we sum over the production of charged light 
hadrons: $\pi^{\pm}, K^{\pm}, p,\bar p$. 

In order to find the values of the BLM scales, we introduce the ratios of the 
BLM to the ``natural'' scale suggested by the kinematic of the process, 
$\mu_N=\sqrt{|\vec k_{1}||\vec k_{2}|}$, so that $m_R=\mu_R^{\rm BLM}/\mu_N$, 
and look for the values of $m_R$ such that Eq.~(\ref{beta0}) is satisfied. 

\begin{figure}[t]
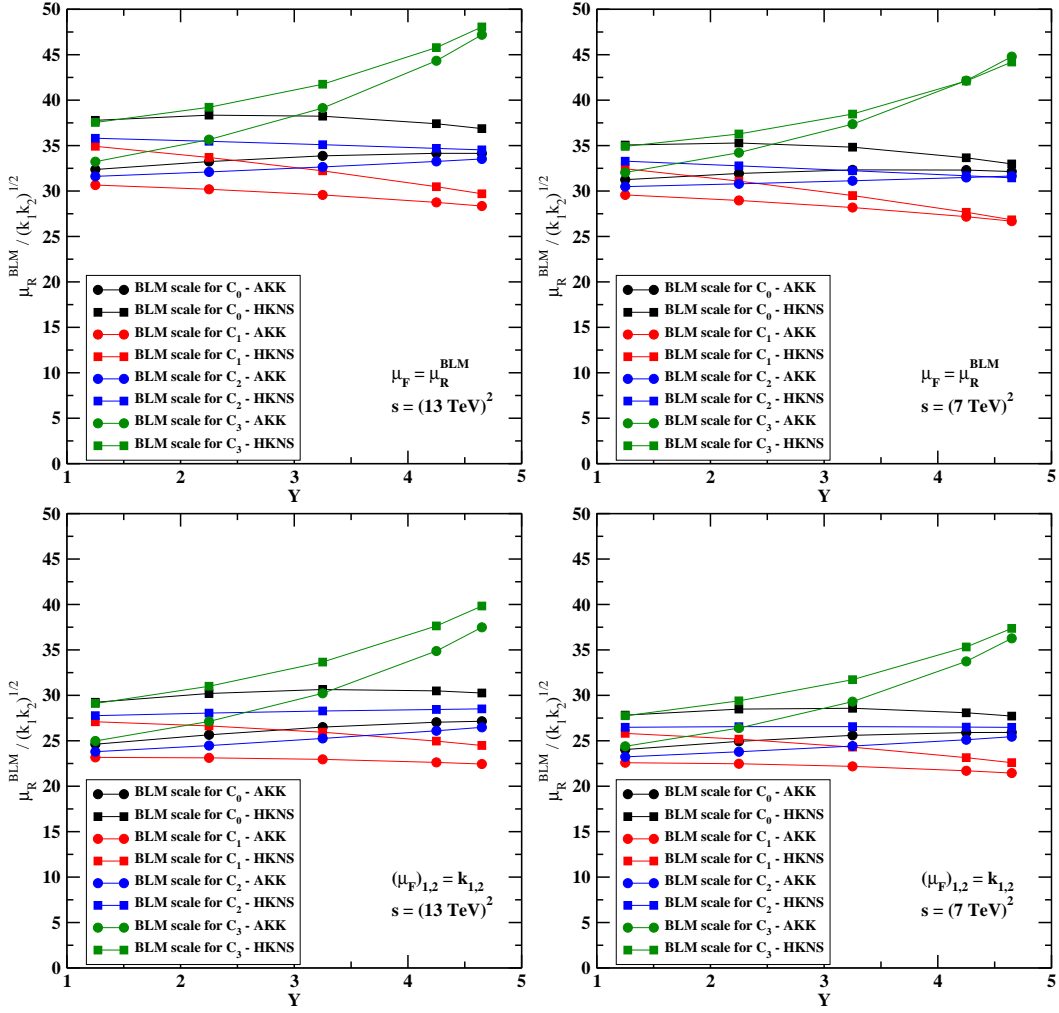

 \centering
 \includegraphics[scale=0.4,clip]{dh-bb-BLM_scales_13TeV.eps}
 \includegraphics[scale=0.4,clip]{dh-bb-BLM_scales_7TeV.eps}
 \includegraphics[scale=0.4,clip]{dh-kb-BLM_scales_13TeV.eps}
 \includegraphics[scale=0.4,clip]{dh-kb-BLM_scales_7TeV.eps}
  \caption[]
{BLM scales for the dihadron production process versus the rapidity
interval $Y$ for $C_n$, $n$=0, 1, 2, 3, for the center-of-mass energies
$\sqrt{s}$=7 and 13 TeV. The top plots are for the choice $\mu_F=\mu_R^{\rm BLM}$,
while the lower ones for $(\mu_F)_{1,2}=|\vec k_{1,2}|$.}
 \label{BLM}
 \end{figure}

The values for $m_R$ are not affected by the inclusion of the NLO corrections
to the impact factor, therefore, for the region $Y\leq 4.8$ and for the
case $\mu_F=\mu_R^{\rm BLM}$, are exactly the same shown in Fig.~1 of
Ref.~\cite{Celiberto:2016hae}; in the case $(\mu_F)_{1,2}=|\vec k_{1,2}|$ they
turn to be generally lower than in the previous case (see Figure~\ref{BLM}
for the summary of all determinations for $m_R$ in the region $Y\leq 4.8$).
In the rapidity region $4.8 < Y \leq 9.4$ we got values for $m_R$ similar
to those shown in Fig.~\ref{BLM}, except for $n=3$, where $m_R$ turned to
be four to five times larger than in the region $Y\leq 4.8$.

Then we plug these scales into our expression for the integrated coefficients
in the BLM scheme (for the derivation see~\cite{Caporale:2015uva}):
\beq{}
\label{eq}
C_n =
\int_{y_{1,\rm min}}^{y_{1,\rm max}}dy_1
\int_{y_{2,\rm min}}^{y_{2,\rm max}}dy_2\int_{k_{1,\rm min}}^{\infty}dk_1
\int_{k_{2,\rm min}}^{\infty}dk_2
\,
\int\limits^{\infty}_{-\infty} d\nu 
\eeq
\beq \nonumber
\frac{e^Y}{s}\,
 e^{Y \bar \alpha^{\rm MOM}_s(\mu^{\rm BLM}_R)\left[\chi(n,\nu)
+\bar \alpha^{\rm MOM}_s(\mu^{\rm BLM}_R)\left(\bar \chi(n,\nu) +\frac{T^{\rm conf}}
{3}\chi(n,\nu)\right)\right]}
\eeq
\[
\times \left(\alpha^{\rm MOM}_s (\mu^{\rm BLM}_R)\right)^2 c_1(n,\nu)c_2(n,\nu)
\left[1+\bar \alpha^{\rm MOM}_s(\mu^{\rm BLM}_R)\left\{\frac{\bar c^{(1)}_1(n,\nu)}
{c_1(n,\nu)}+\frac{\bar c^{(1)}_2(n,\nu)}{c_2(n,\nu)}+\frac{2T^{\rm conf}}{3}
\right\} \right] \, .
\]
The coefficient $C_0$ gives the total cross sections and the ratios
$C_n/C_0 = \langle\cos(n\phi)\rangle$ determine the values of the mean cosines,
or azimuthal correlations, of the produced hadrons. In Eq.~(\ref{eq}), 
$\bar \chi(n,\nu)$ is the eigenvalue of NLA BFKL kernel~\cite{Kotikov:2000pm}
 and its expression is given, {\it e.g.} in Eq.~(23) 
of~\cite{Caporale:2012ih}, whereas $\bar c^{(1)}_{1,2}$ are the NLA parts of 
the hadron vertices~\cite{hadrons}. 

As anticipated, we give predictions for $C_n$ by fixing the factorization scale
$\mu_F$ in two different ways:
\begin{enumerate}
\item 
$\mu_F = \mu_R = \mu_R^{\rm BLM}$;
\item
$(\mu_F)_{1,2}  = |\vec k_{1,2}|$.
\end{enumerate}

All calculations are done in the MOM scheme. For comparison, we present results 
for the $\phi$-averaged cross section $C_0$ in the $\overline{\rm MS}$ scheme 
(as implemented in Eq.~(\ref{Cm}))
for $\sqrt{s} = 7, 13$ TeV and for $Y \leq 4.8, 9.4$.
In this case, we choose  natural values for $\mu_R$, {\it i.e.} 
$\mu_R = \mu_N = \sqrt{|\vec k_{1}||\vec k_{2}|}$, and the option 2.,
{\it i.e.} $(\mu_F)_{1,2}  = |\vec k_{1,2}|$ for the factorization scale. 

\subsection{Used tools and uncertainty estimation}
\label{numerics}

We performed all numerical calculations in \textsc{Fortran}, 
choosing a two-loop running coupling setup 
with $\alpha_s\left(M_Z\right)=0.11707$ and five quark flavors.
It is known that potential sources of uncertainty 
could be due to the particular PDF and FF parametrizations used. 
For this reason, we did preliminary tests by using 
three different NLO PDF sets, expressly:  
MSTW~2008~\cite{Martin:2009iq}, 
MMHT~2014~\cite{Harland-Lang:2014zoa} 
(which is the successor of the MSTW~2008 one), 
and CT~2014~\cite{Dulat:2015mca}, 
and convolving them with the three following NLO FF routines: 
AKK~\cite{Albino:2008fy}, 
DSS~\cite{DSS}, 
and HNKS~\cite{Hirai:2007cx}. 
Our tests have shown no significant discrepancy 
when different PDF sets are used in our kinematic range. 
In view of this result, in the final calculations 
we selected the MSTW 2008 PDF set (which was successfully used 
in various analyses of inclusive semi-hard processes at LHC, 
including our previous studies of Mueller-Navelet jets), 
together with the FF interfaces mentioned above.
We do not show the results with the DSS routine, 
since they would be hardly distinguishable from those with 
the HKNS parametrization.

Specific \textsc{CERN} program libraries~\cite{cernlib} 
were used to evaluate the azimuthal coefficients given in Eq.~(\ref{eq}), 
which requires a complicated 8-dimensional numerical integration 
(the expressions for $\bar c^{(1)}_{1,2}$ contain 
an additional longitudinal fraction integral in 
comparison to the formulas for the LLA vertices, given in Eqs.~(\ref{c1}) 
and~(\ref{c2})).
Furthermore, slightly modified versions of the \cod{Chyp}~\cite{chyp} 
and \cod{Psi}~\cite{rpsi} routines were used to calculate 
the Gauss hypergeometric function $_2F_1$ 
and the real part of the $\psi$ function, respectively.

The most significant uncertainty comes from the numerical 4-dimensional
integration over the two transverse momenta $|\vec k_{1,2}|$, 
the rapidity $y_1$, and over $\nu$.  
Its effect was directly estimated by \cod{Dadmul} integration
routine~\cite{cernlib}.
The other three sources of uncertainty, which are respectively:
the one-dimensional integration over the parton fraction $x$
needed to perform the convolution between PDFs and FFs 
in the LO/NLO impact factors (see Eq.~(\ref{c1})~and~(\ref{c11})),
the one-dimensional integration over the longitudinal momentum fraction
$\zeta$ in the NLO impact factor correction (see Eqs.~(\ref{c11})),
and the upper cutoff in the numerical integrations over
$|\vec k_{1,2}|$ and $\nu$, are negligible with respect to
the first one. For this reason the error bars 
of all predictions presented in this work are just those given by the
\cod{Dadmul} routine.

\subsection{Discussion}
\label{discussion}

\begin{figure}[t]
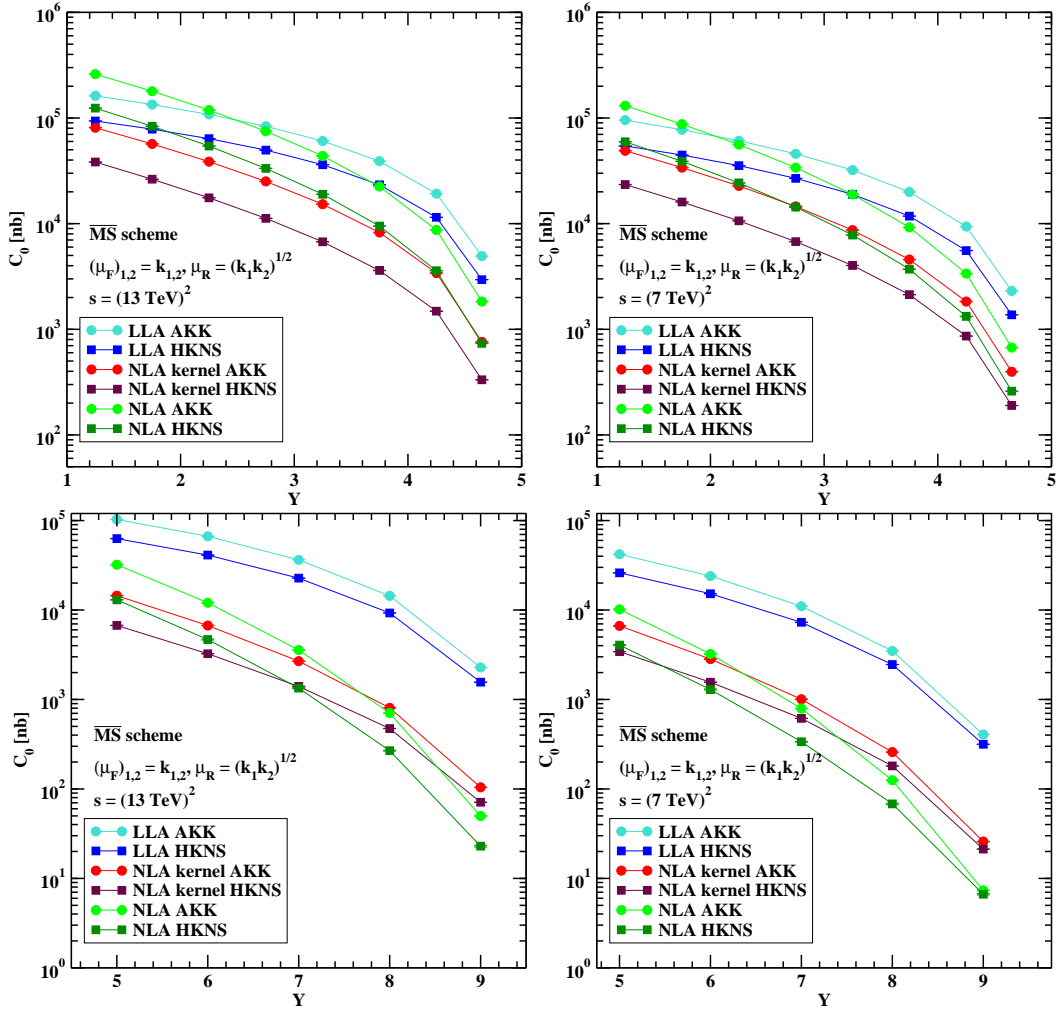

\centering
   \includegraphics[scale=0.40,clip]{C0_MSb_ns_13TeV.eps}
   \includegraphics[scale=0.40,clip]{C0_MSb_ns_7TeV.eps}

   \includegraphics[scale=0.40,clip]{C0_MSb_ns_LY_13TeV.eps}
   \includegraphics[scale=0.40,clip]{C0_MSb_ns_LY_7TeV.eps}

   \caption{$Y$-dependence of $C_0$ in the $\overline{\rm MS}$ scheme 
(as implemented in Eq.~(\ref{Cm})) at natural scales for
$\mu_R$ and $\mu_F$, $\sqrt{s} = 7, 13$ TeV, and in the two cases 
  of $Y \leq 4.8$ and $Y \leq 9.4$.  Here and in the following figure
  captions ``LLA'' means pure leading logarithmic approximation, ``NLA kernel''
  means inclusion of the NLA corrections from the kernel only, ``NLA''
  stands for full inclusion of NLA corrections, {\it i.e.} both from
  the kernel and the hadron vertices.}
\label{fig:C0MSbNS}
\end{figure}

In Fig.~\ref{fig:C0MSbNS} we present our results for $C_0$ in the
$\overline{\rm MS}$ scheme (as implemented in Eq.~(\ref{Cm})) for 
 we already specified above the scale settings  $\sqrt{s} = 7, 13$ TeV,
and in the two cases of $Y \leq 4.8$ and $Y \leq 9.4$.
We clearly see that NLA corrections become negative 
with respect to the LLA prediction when $Y$ grows.
Besides, it is interesting to note that the full NLA approach predicts larger
values for the cross sections in comparison to the case where only NLA
corrections to the BFKL kernel are taken into account. It means that the
inclusion into the analysis of the NLA corrections to the hadron vertices
makes the predictions for the cross sections somewhat bigger and parially
compensates the large negative effect from the NLA corrections to the BFKL
kernel.

\begin{figure}[p]
\centering

   \includegraphics[scale=0.40,clip]{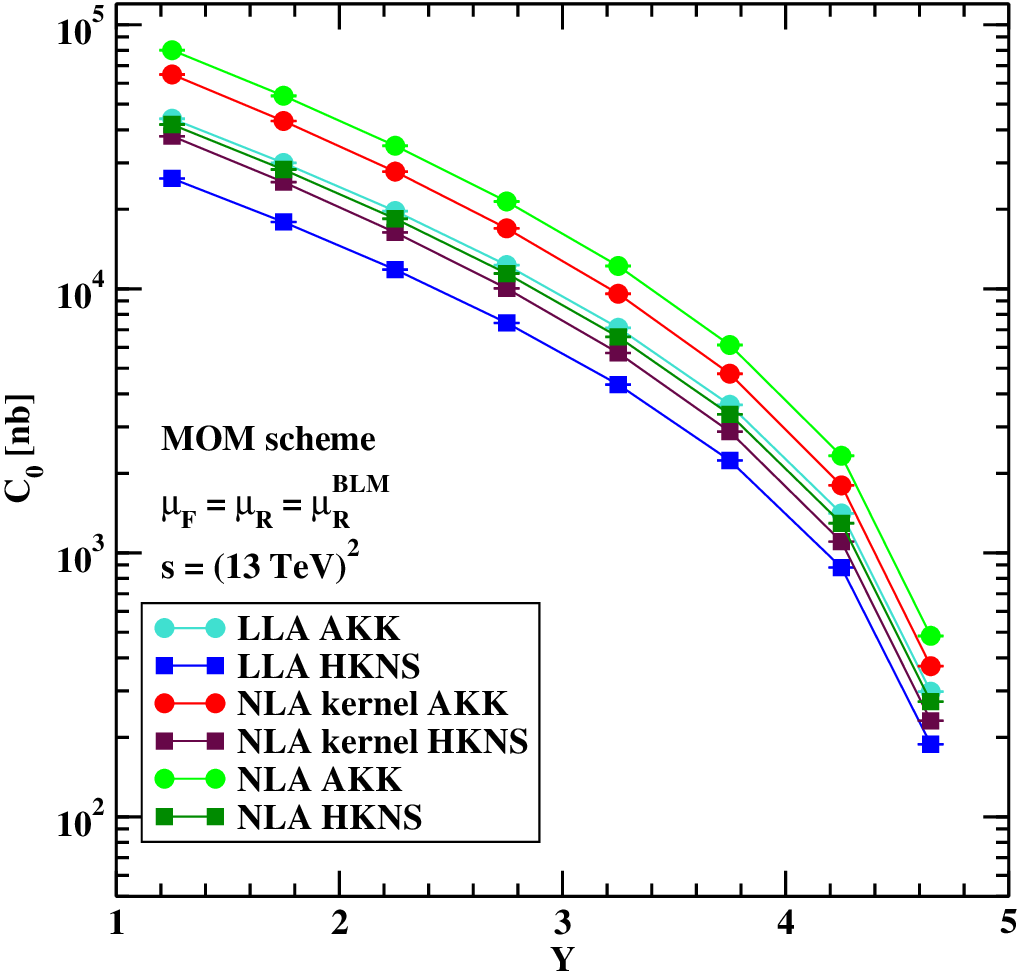}
   \includegraphics[scale=0.40,clip]{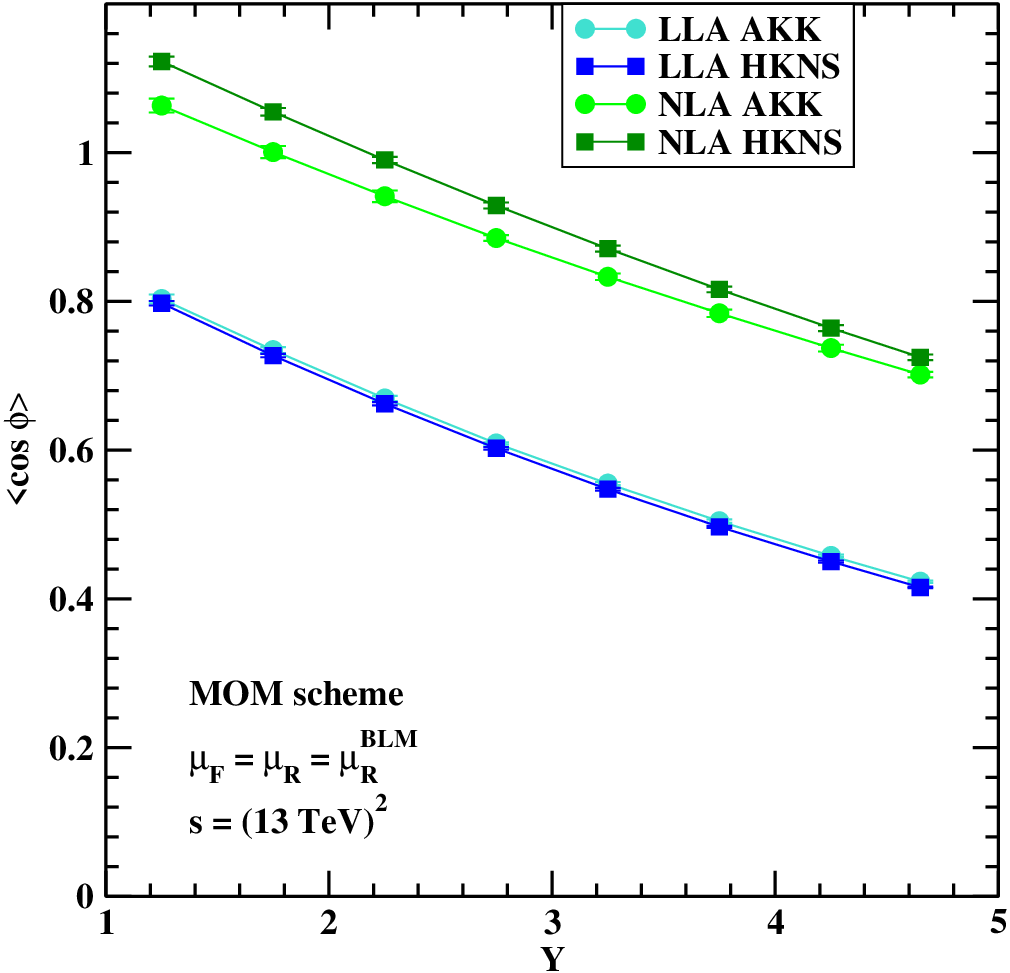}

   \includegraphics[scale=0.40,clip]{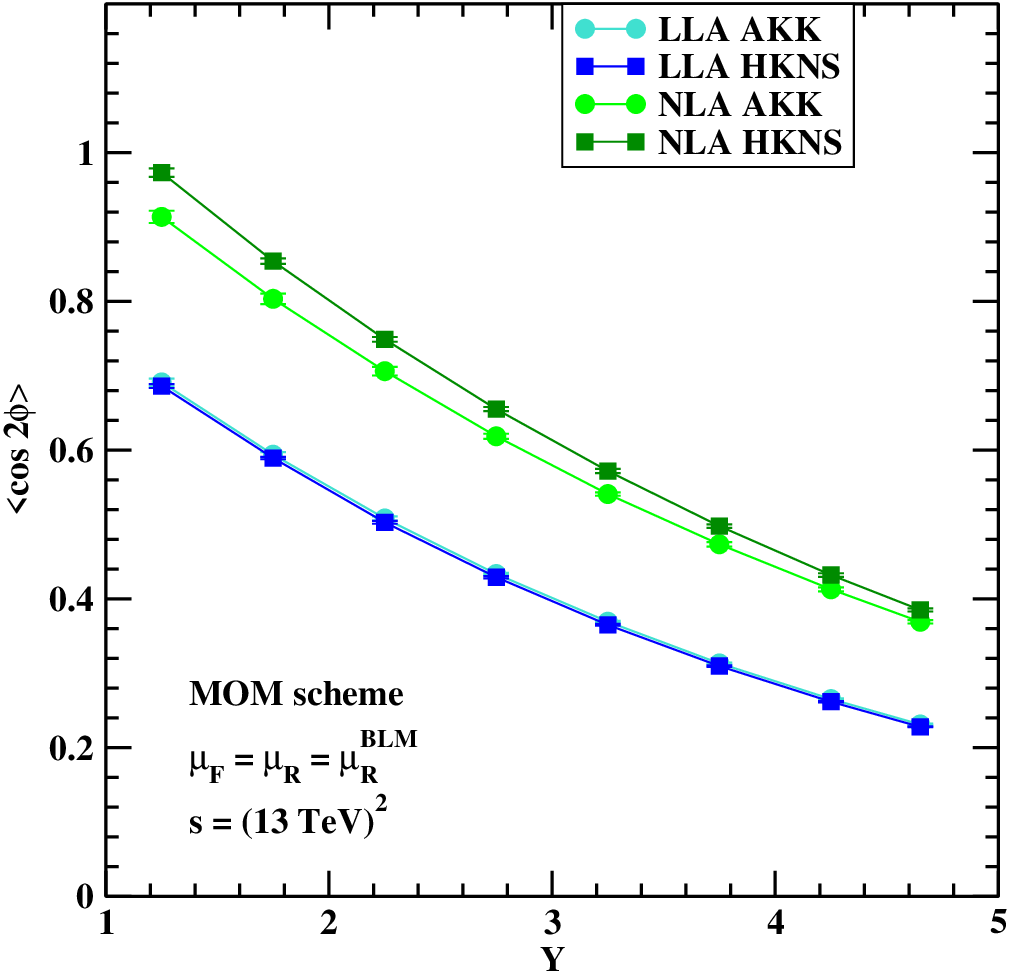}
   \includegraphics[scale=0.40,clip]{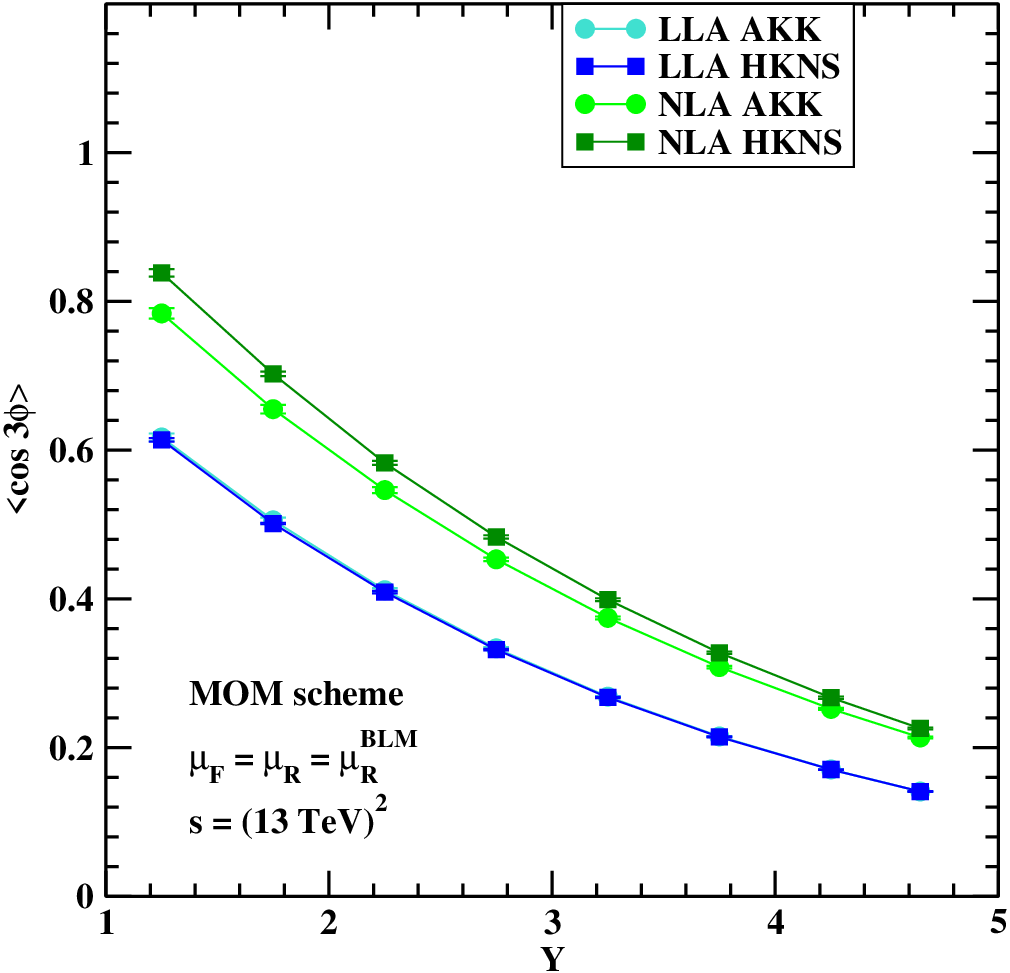}

   \includegraphics[scale=0.40,clip]{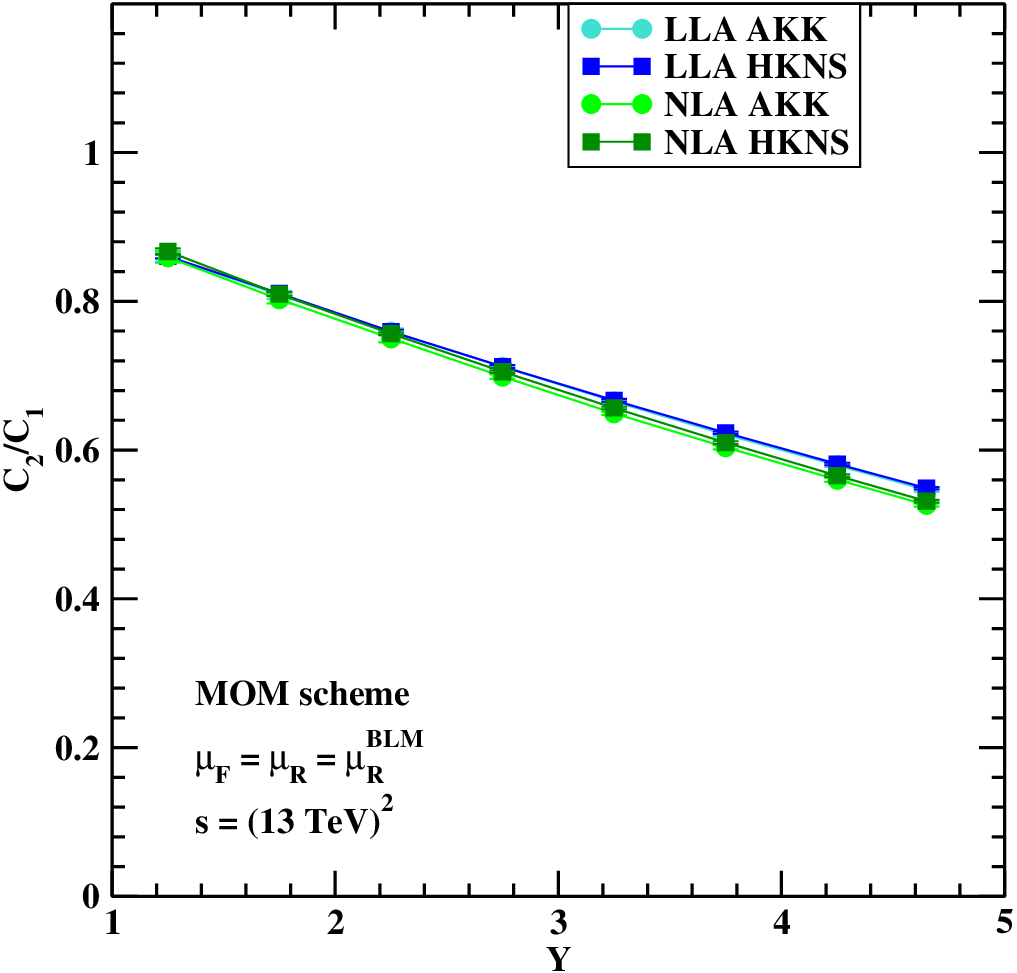}
   \includegraphics[scale=0.40,clip]{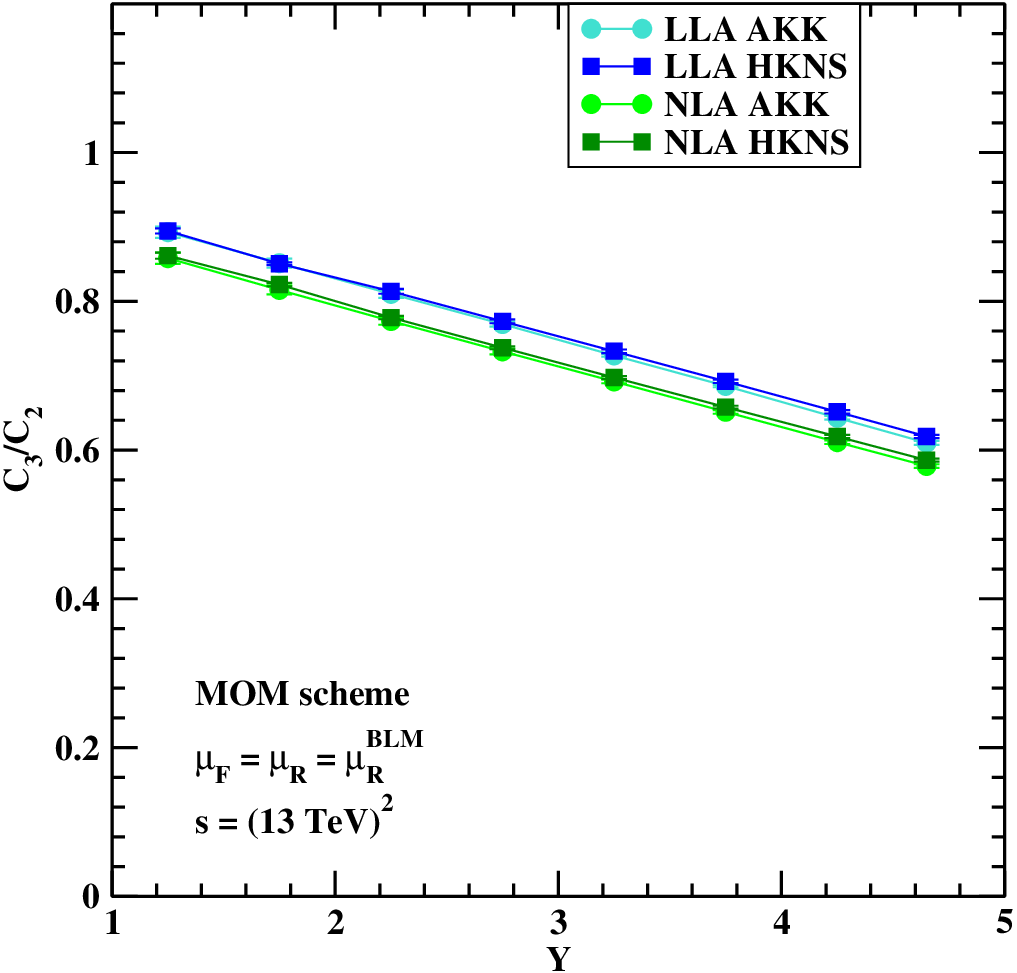}
\caption{$Y$-dependence of $C_0$ and of several ratios $C_m/C_n$ for 
$\mu_F = \mu^{\rm BLM}_R$, $\sqrt{s} = 13$ TeV, and $Y \leq 4.8$.}
\label{fig:blm13}
\end{figure}

\begin{figure}[p]
\centering

   \includegraphics[scale=0.40,clip]{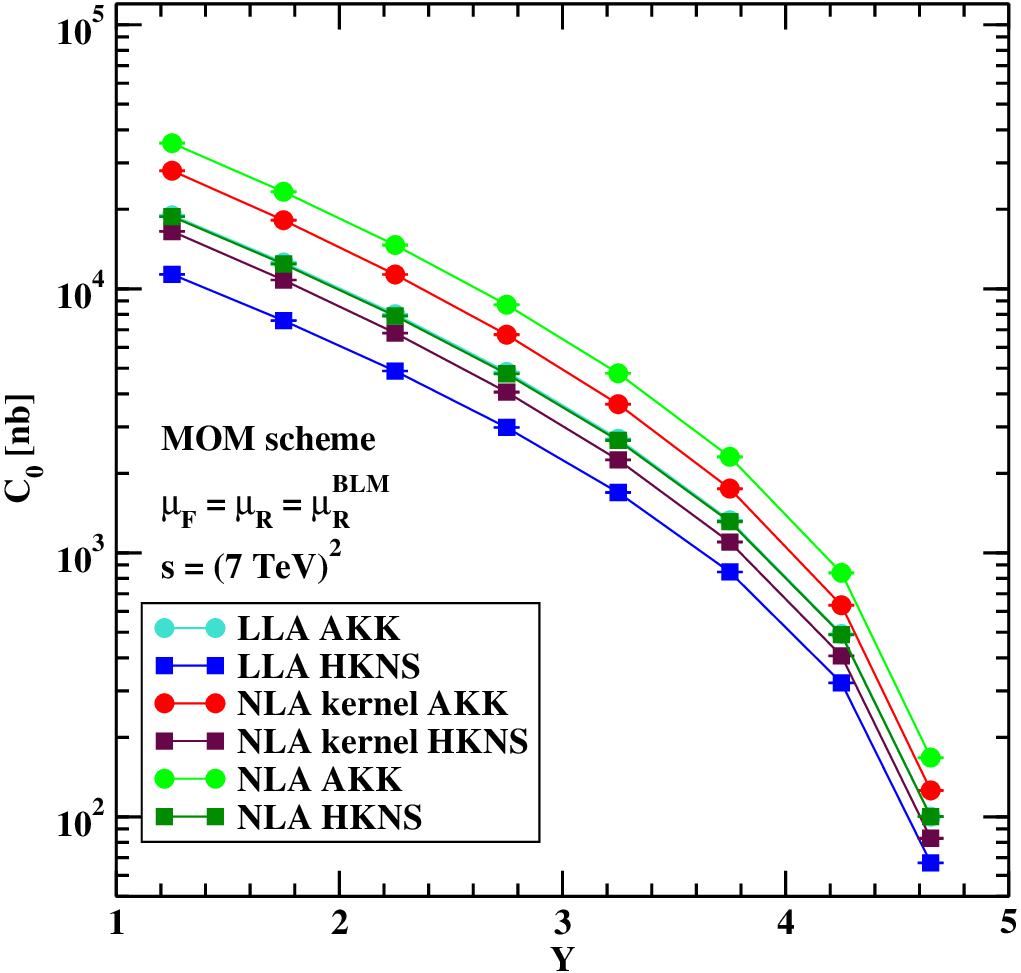}
   \includegraphics[scale=0.40,clip]{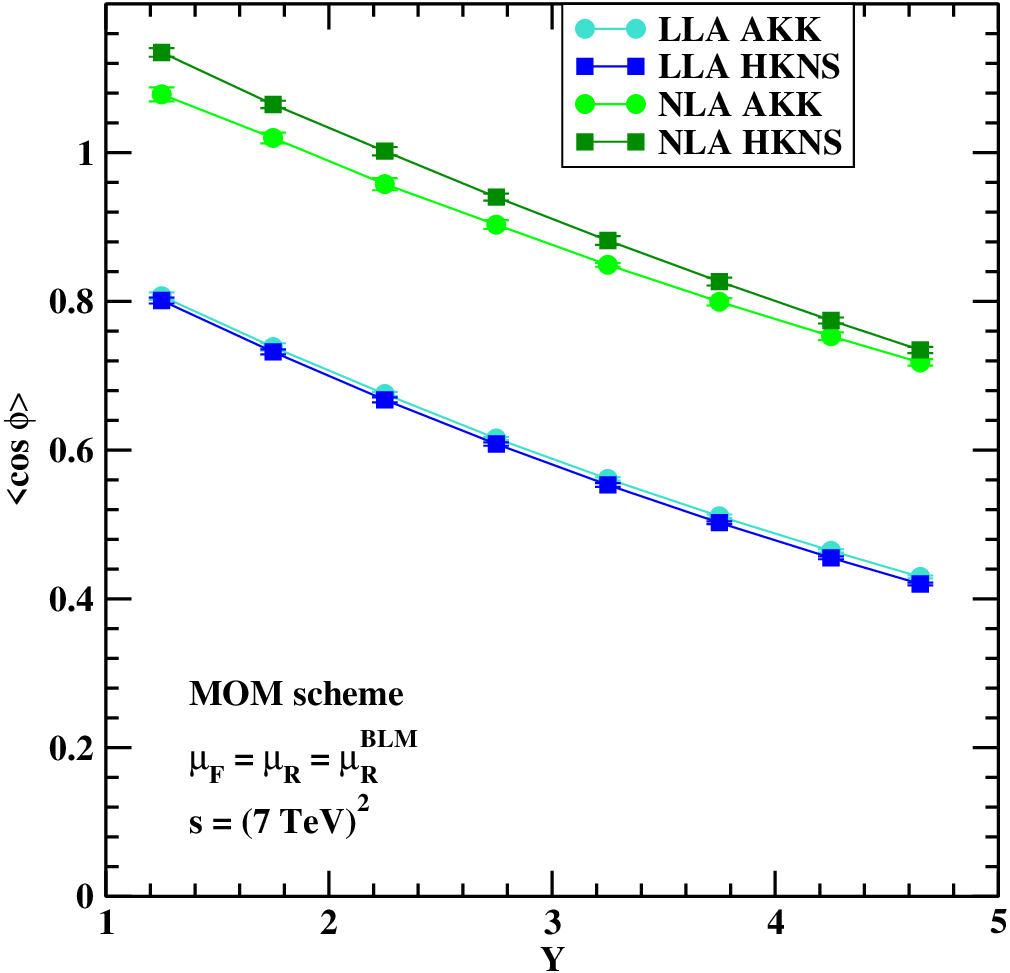}

   \includegraphics[scale=0.40,clip]{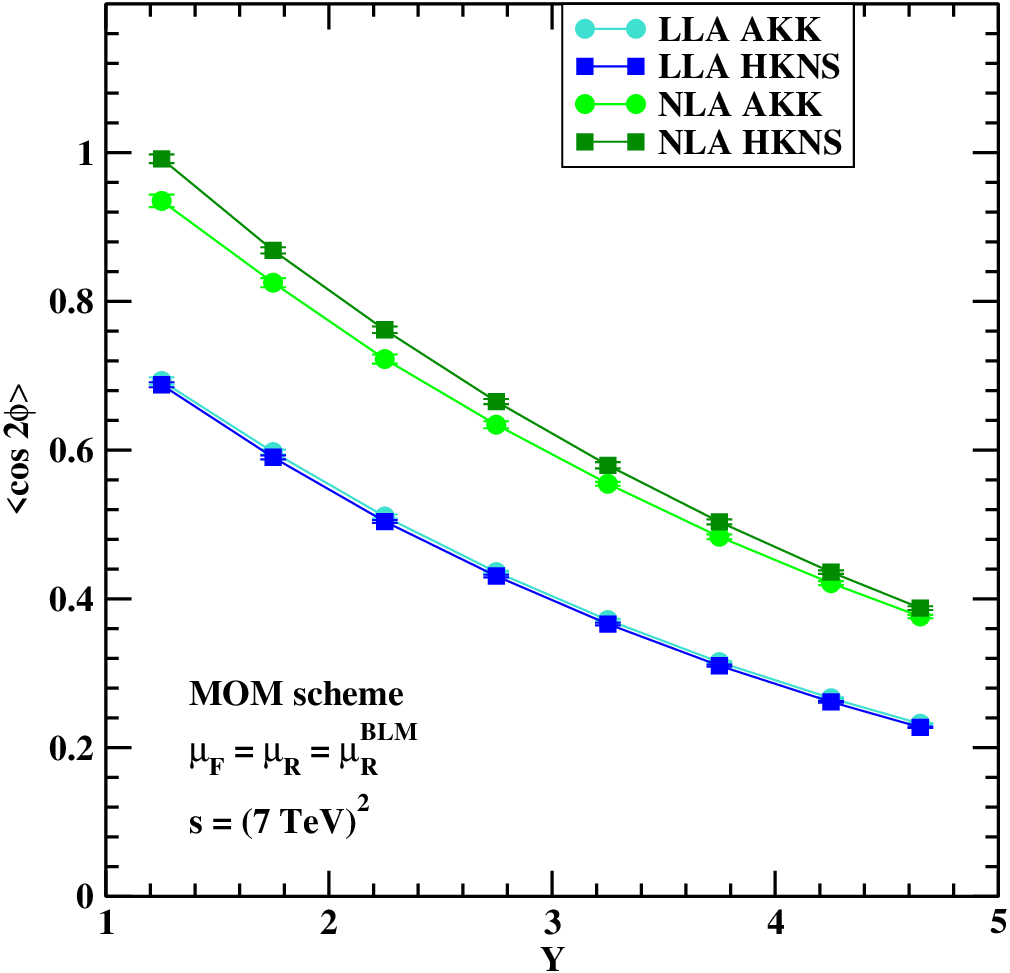}
   \includegraphics[scale=0.40,clip]{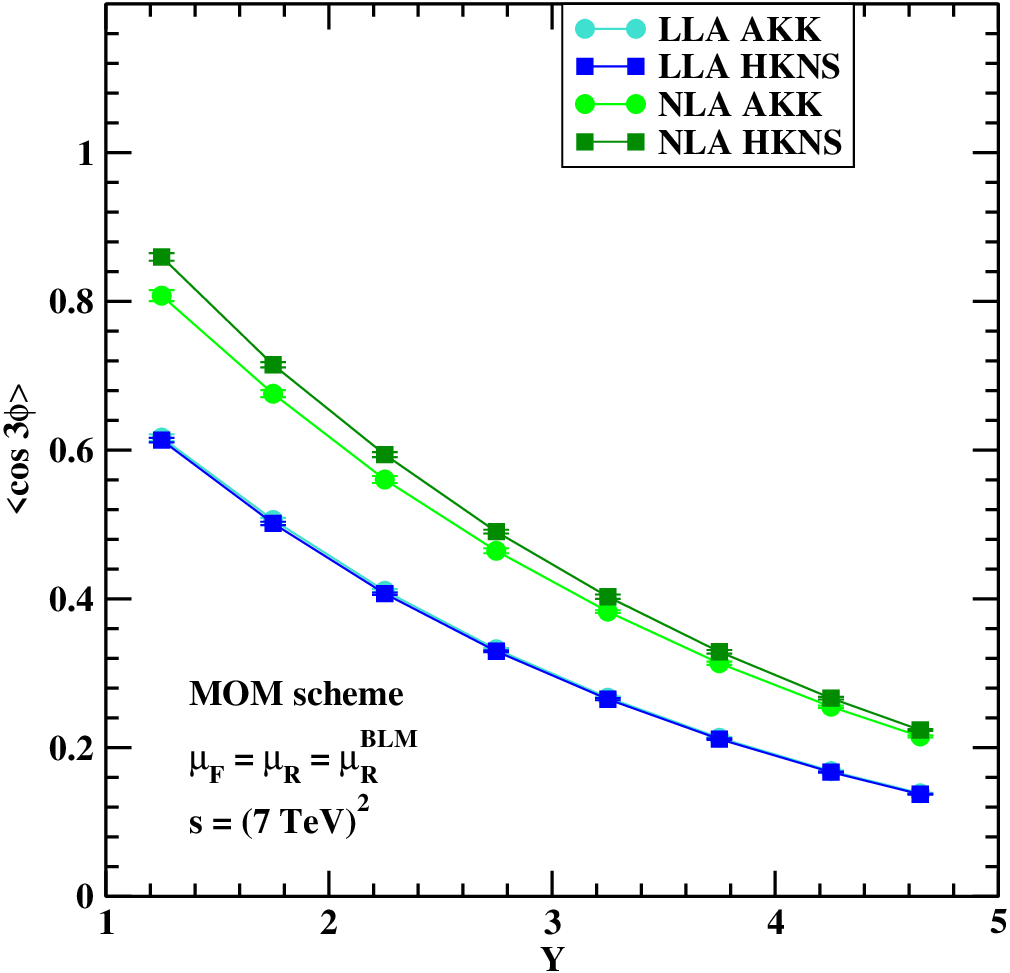}

   \includegraphics[scale=0.40,clip]{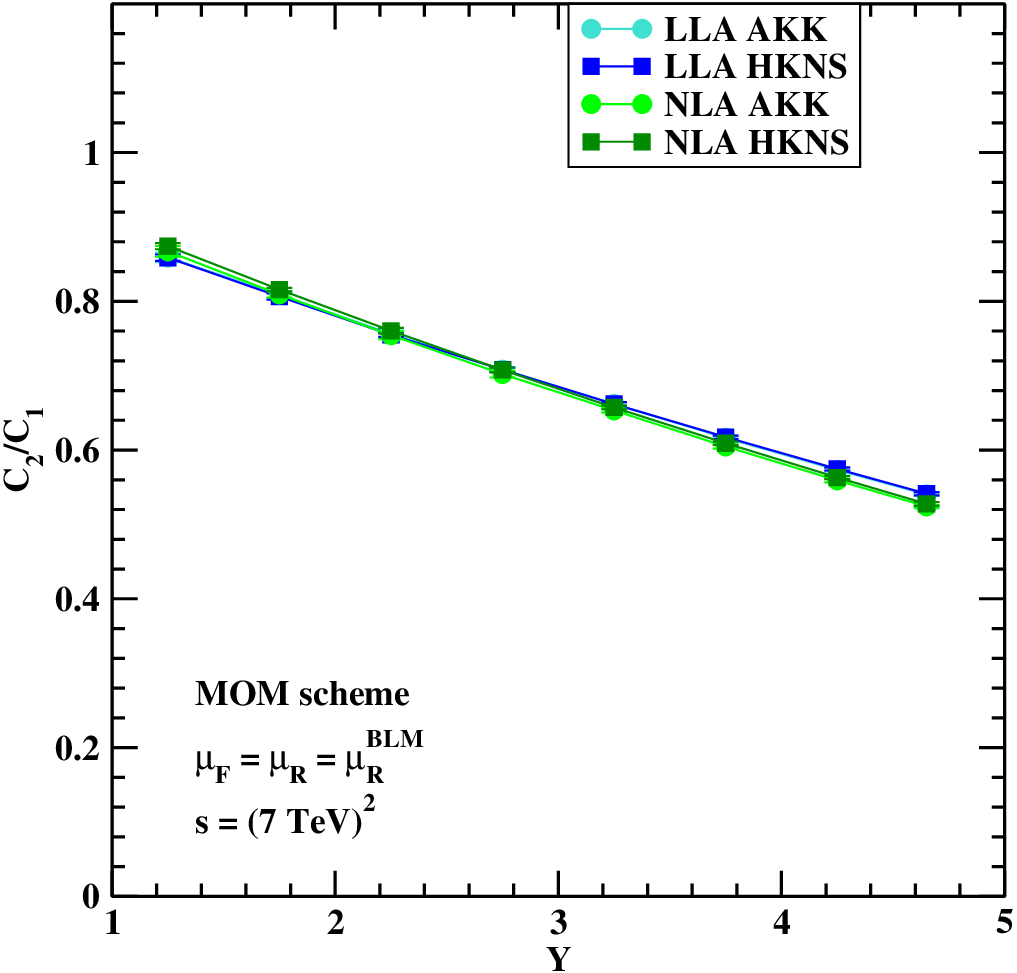}
   \includegraphics[scale=0.40,clip]{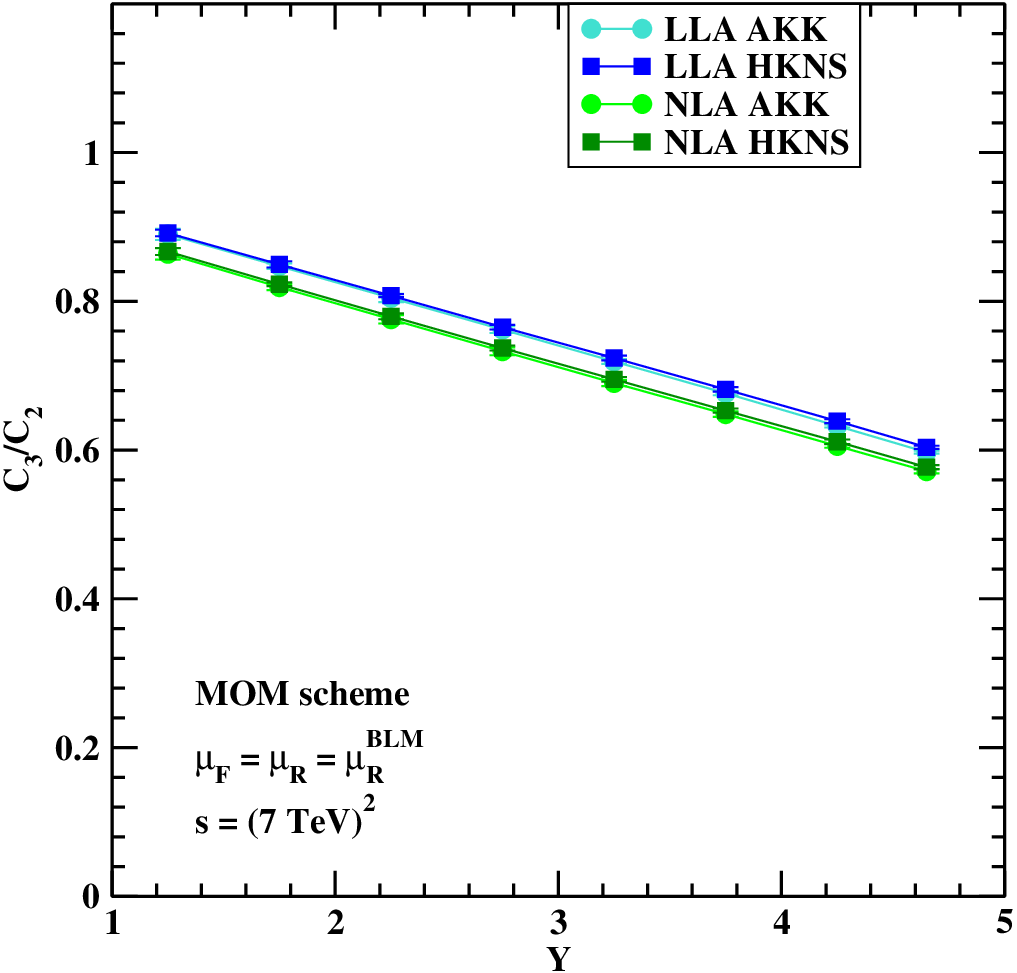}
\caption{$Y$-dependence of $C_0$ and of several ratios $C_m/C_n$ for 
$\mu_F = \mu^{\rm BLM}_R$, $\sqrt{s} = 7$ TeV, and $Y \leq 4.8$.}
\label{fig:blm7}
\end{figure}

The other results we presented below are obtained using BLM in the MOM scheme,
as it is given in Eq.~(\ref{eq}). 
In Figs.~\ref{fig:blm13}~and~\ref{fig:blm7} we present our results 
for $C_0$ and for several ratios $C_m/C_n$ at $\sqrt{s}=13$ and $7$ TeV,
respectively; $\mu_F$ is set equal to $\mu^{\rm BLM}_R$,
while $Y \leq 4.8$. It is worth to note that in this case the NLA corrections 
to $C_0$ are positive, so they increase the value 
of the $\phi$-averaged cross section at all values of $Y$. 
This is the result of the combination of two distinct effects:
on one side, we already saw in Ref.~\cite{Celiberto:2016hae} 
that changing the renormalization scheme produces 
a non-exponentiated extra factor in Eq.~(\ref{eq}) proportional
to $T^{\rm conf}$, and that is positive. On the other side, we found that 
the $C_{gg}$ coefficient in Eq.~(\ref{c11})
gives a large and positive contribution to the NLO impact factor. 
We see also that NLA corrections increase the azimuthal correlations: 
$C_1/C_0$, $C_2/C_0$, and $C_3/C_0$, while their effect is small with respect
to LLA predictions in their ratios, $C_2/C_1$ and $C_3/C_2$. 
The value of $C_1/C_0$ for $Y \leq 2.75$ in some cases exceeds $1$.
We consider this as an effect due to the fact that, at very small $Y$,
which corresponds to the small values of partonic subenergies $\hat s$, 
we are crossing the applicability limit of the BFKL approach, which
systematically neglects any contributions that are suppressed by the powers of 
$\hat s$.  

\begin{figure}[p]
\centering

   \includegraphics[scale=0.40,clip]{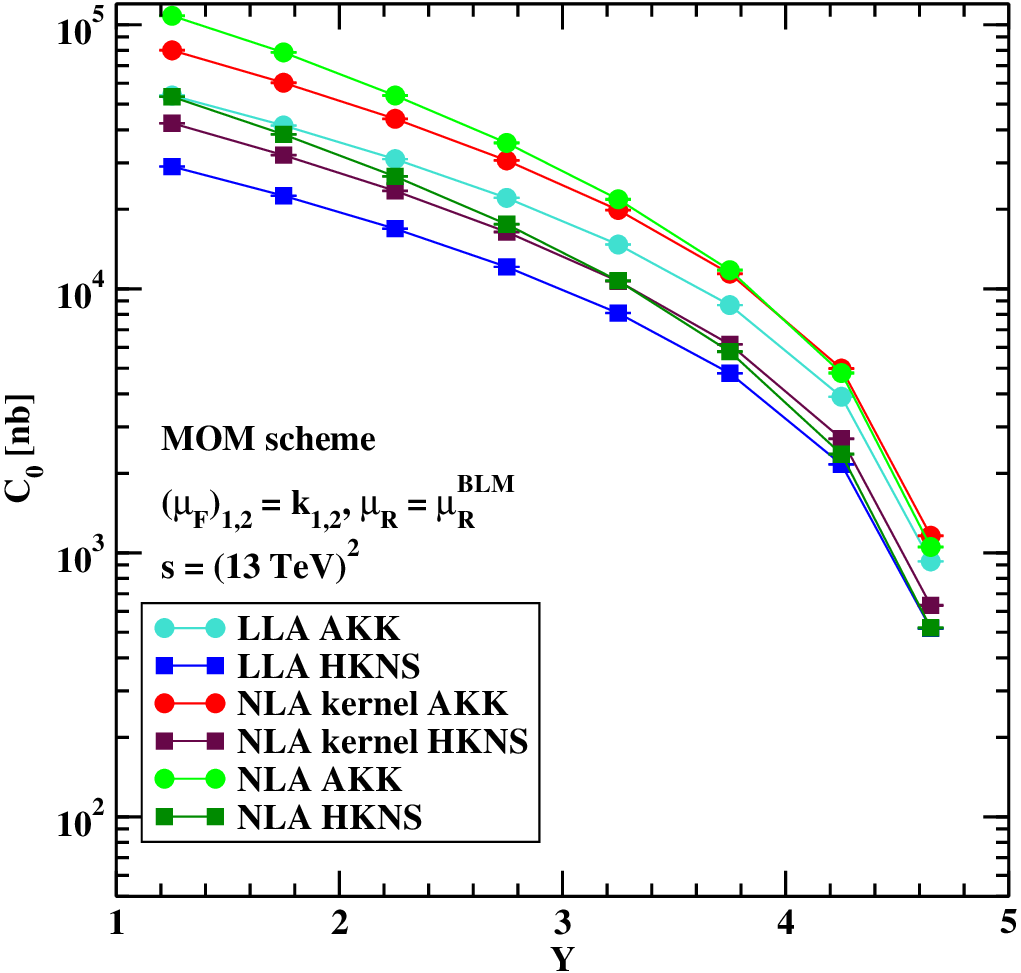}
   \includegraphics[scale=0.40,clip]{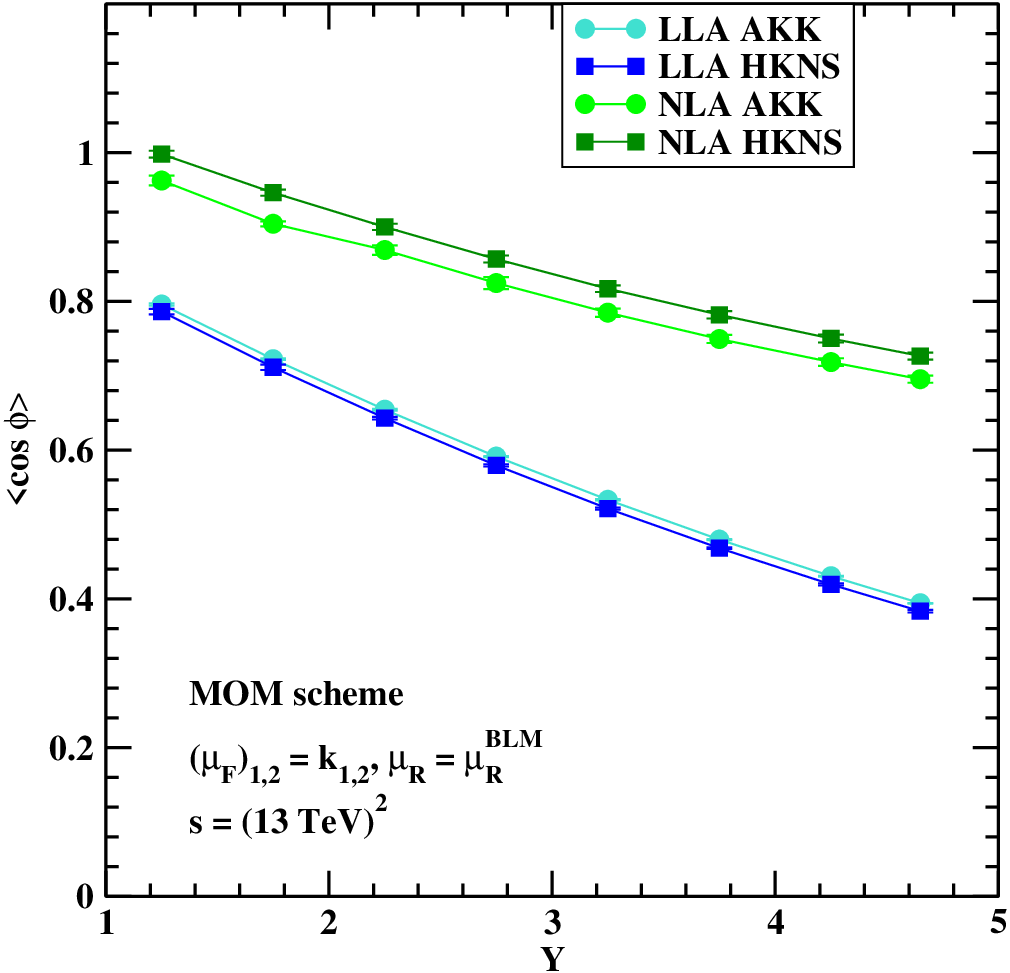}

   \includegraphics[scale=0.40,clip]{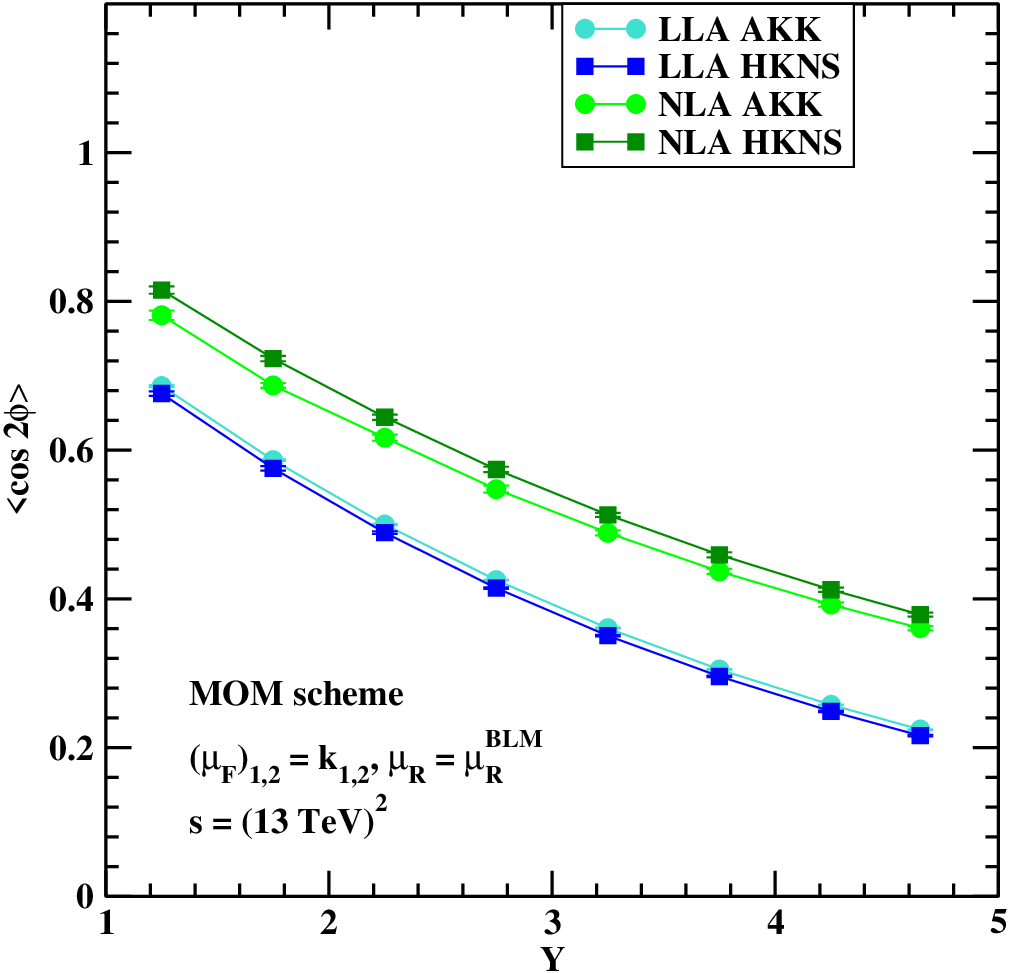}
   \includegraphics[scale=0.40,clip]{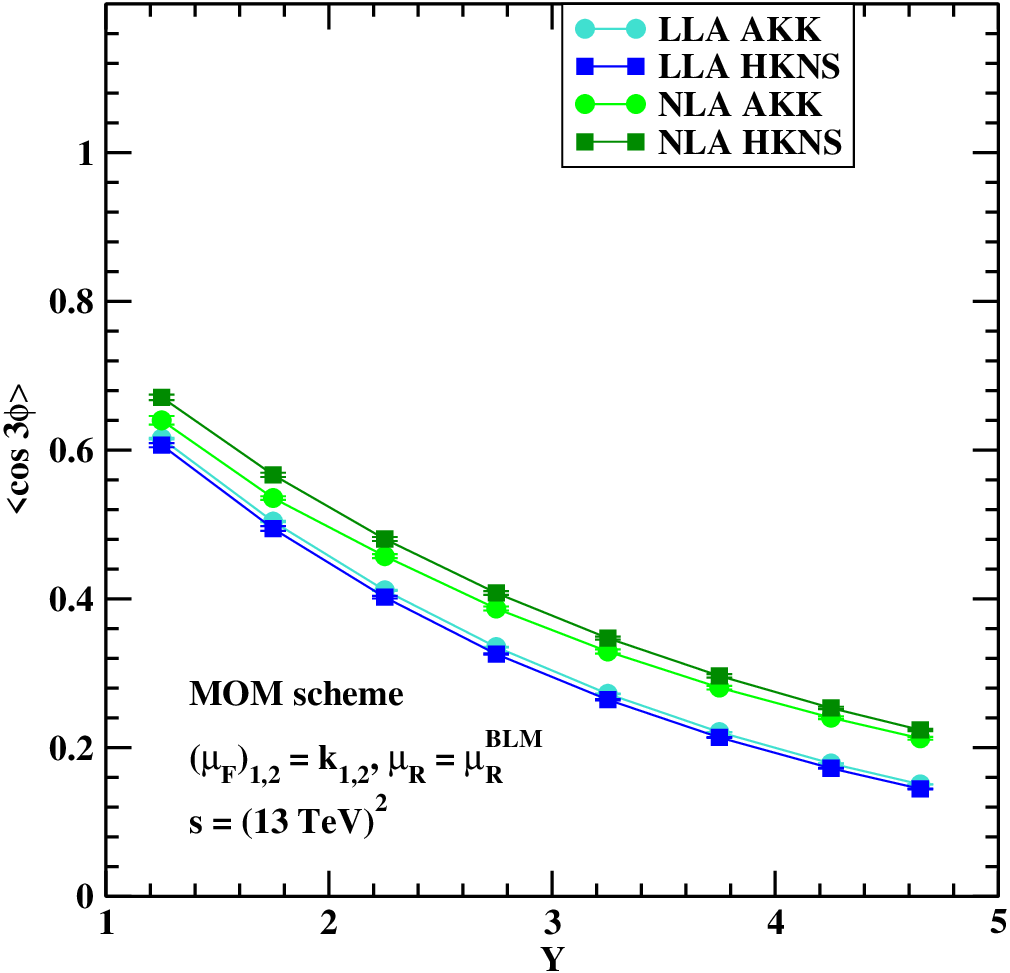}

   \includegraphics[scale=0.40,clip]{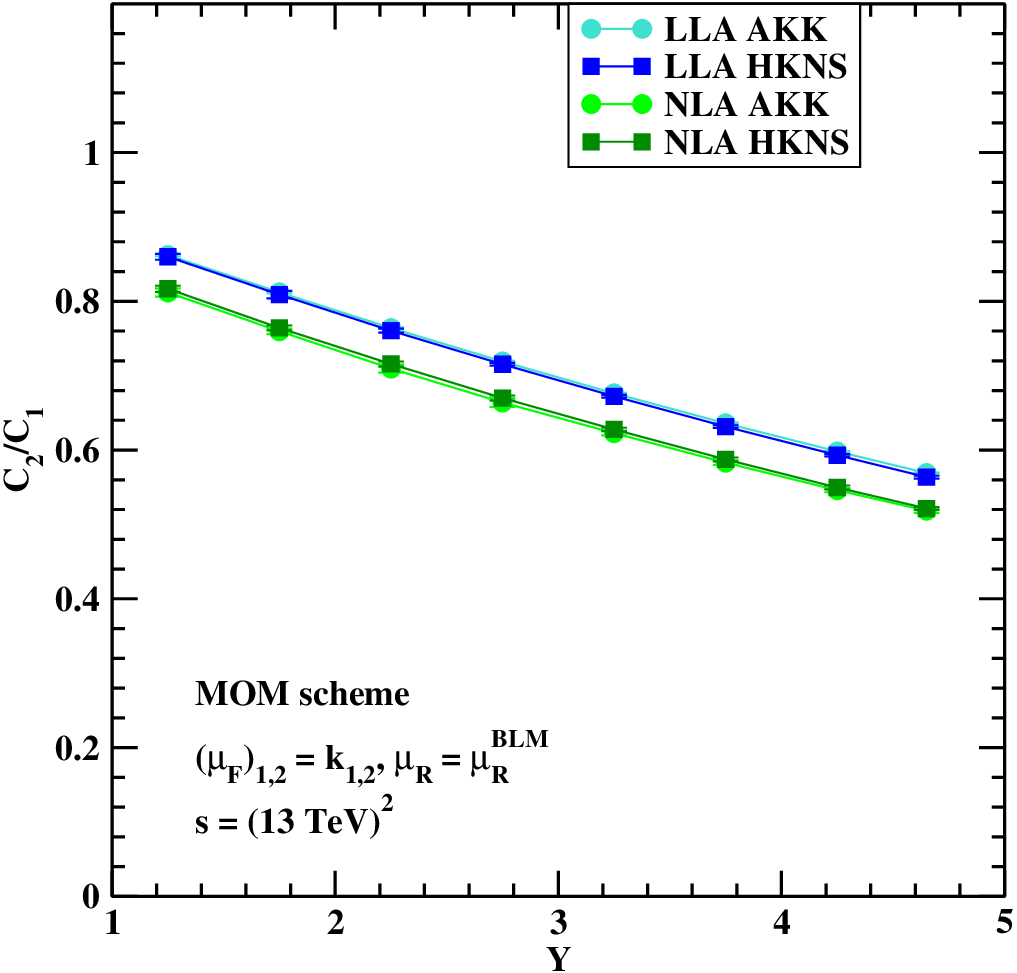}
   \includegraphics[scale=0.40,clip]{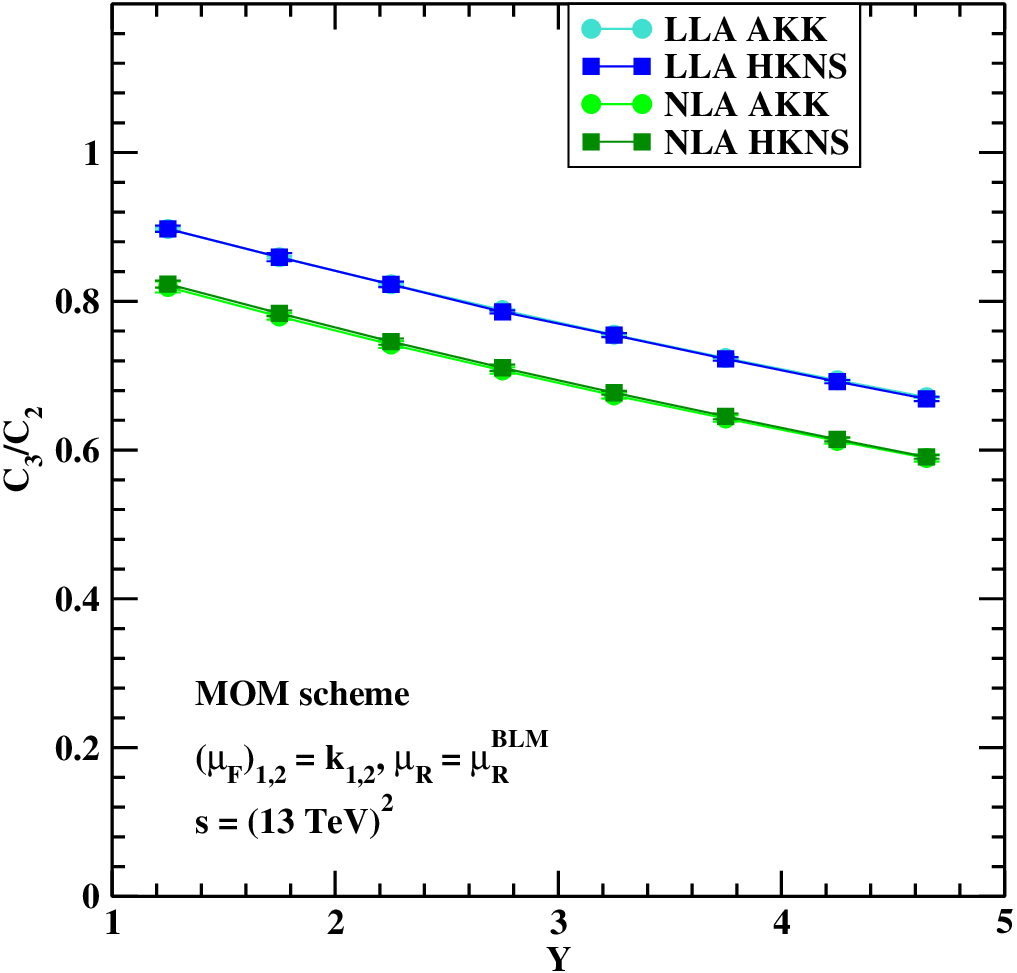}
\caption{$Y$-dependence of $C_0$ and of several ratios $C_m/C_n$ for 
$(\mu_F)_{1,2} = |\vec k_{1,2}|$, $\sqrt{s} = 13$ TeV, and $Y \leq 4.8$.}
\label{fig:ns13}
\end{figure}

\begin{figure}[p]
\centering

   \includegraphics[scale=0.40,clip]{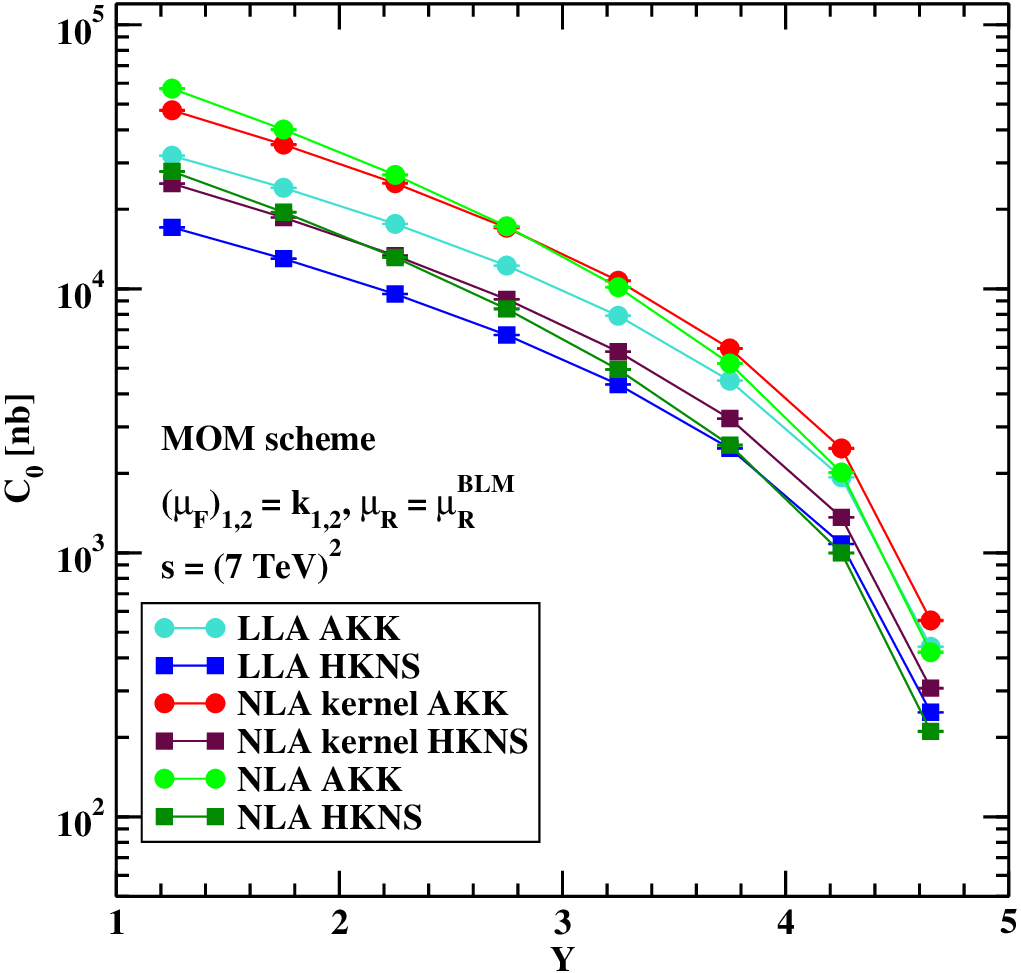}
   \includegraphics[scale=0.40,clip]{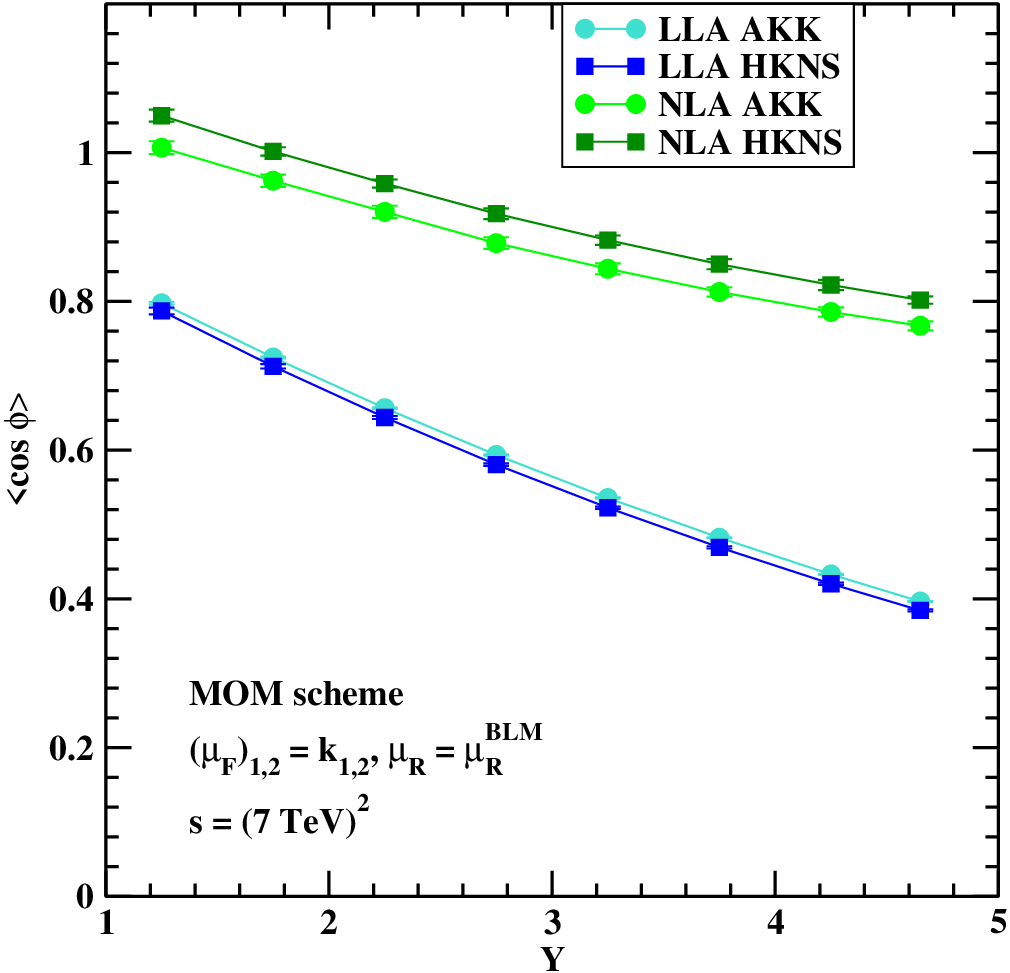}

   \includegraphics[scale=0.40,clip]{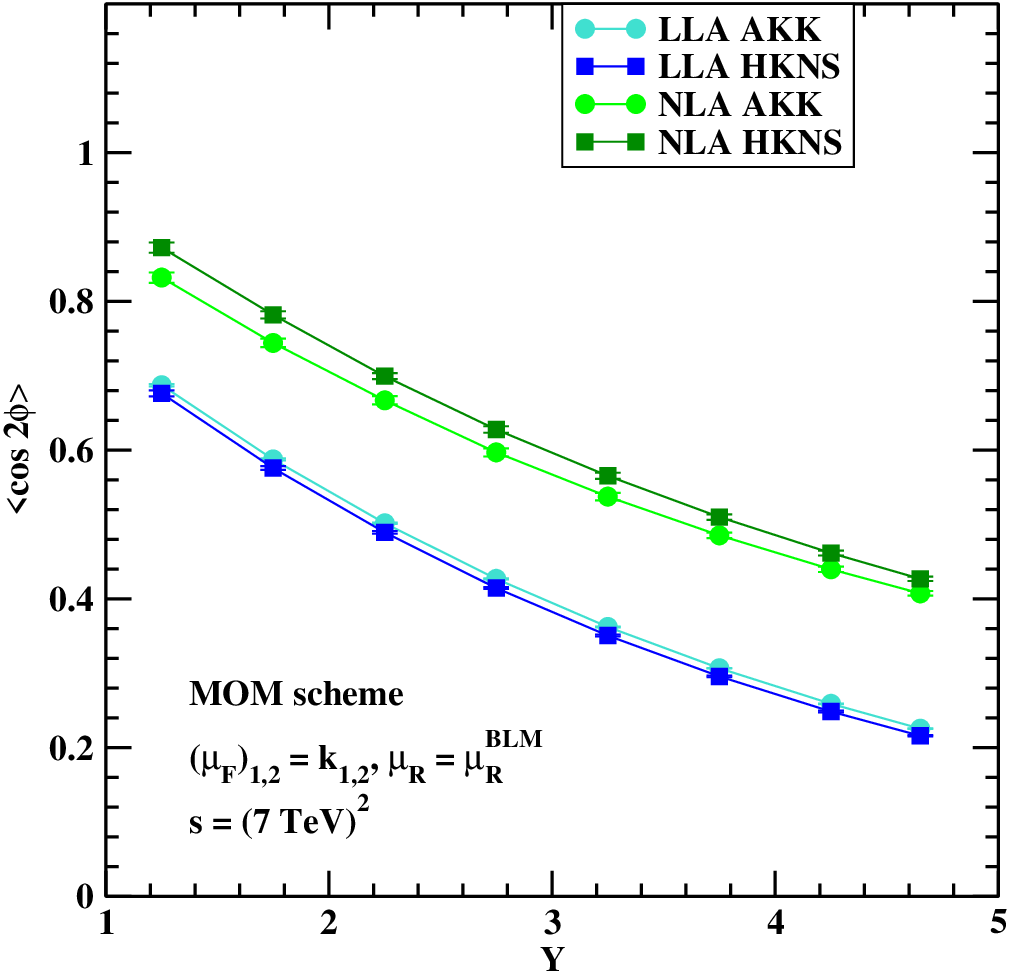}
   \includegraphics[scale=0.40,clip]{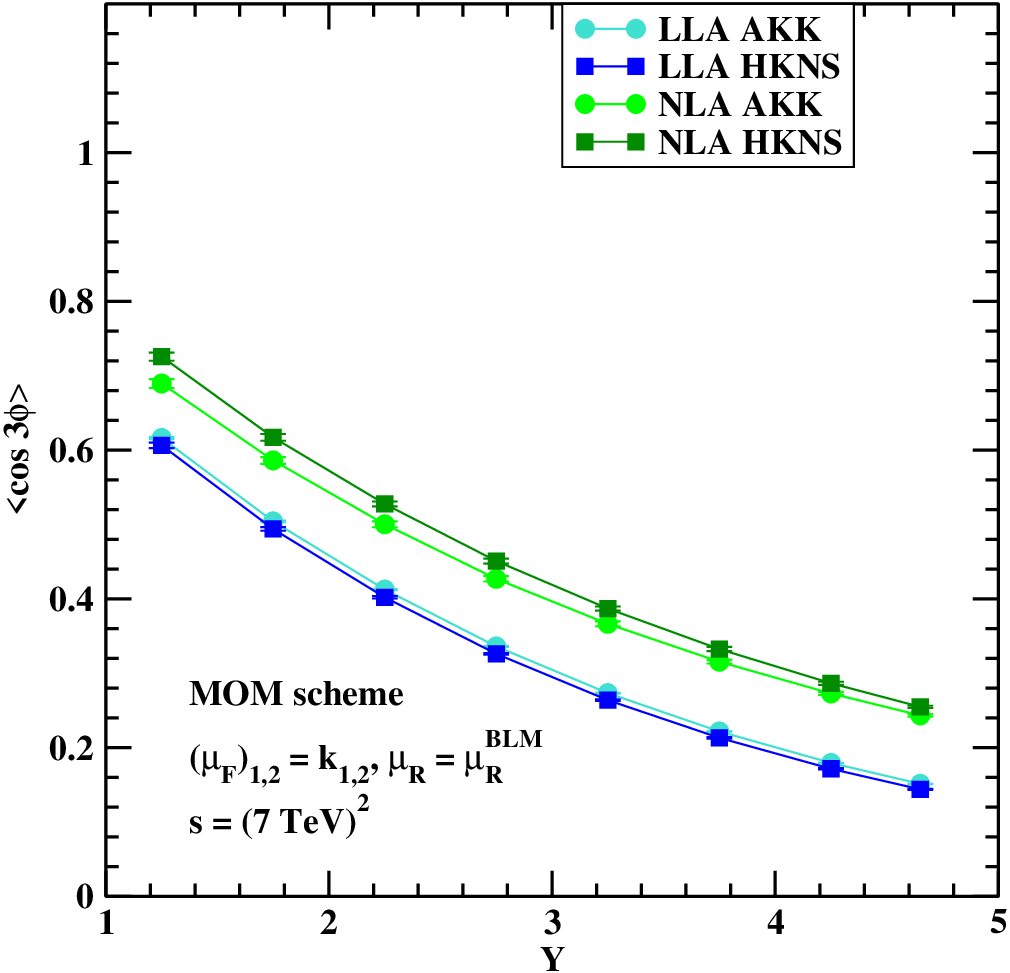}

   \includegraphics[scale=0.40,clip]{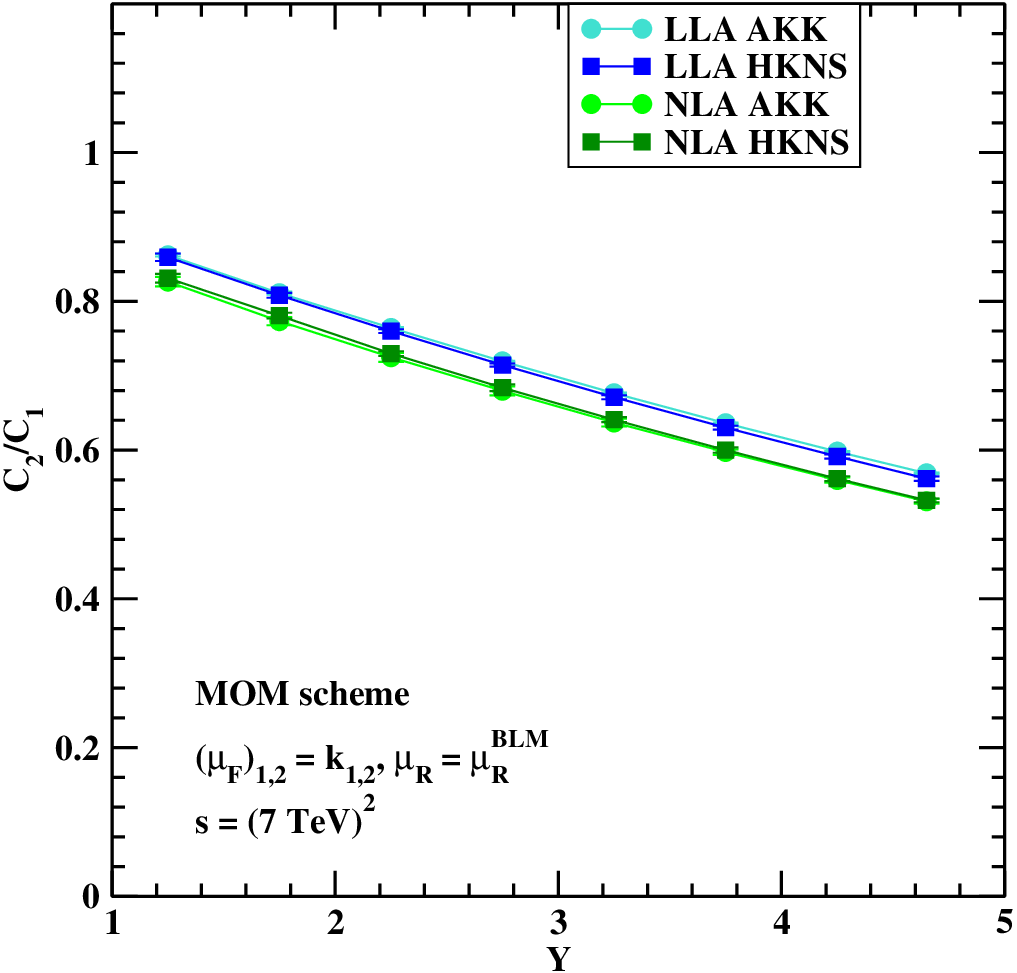}
   \includegraphics[scale=0.40,clip]{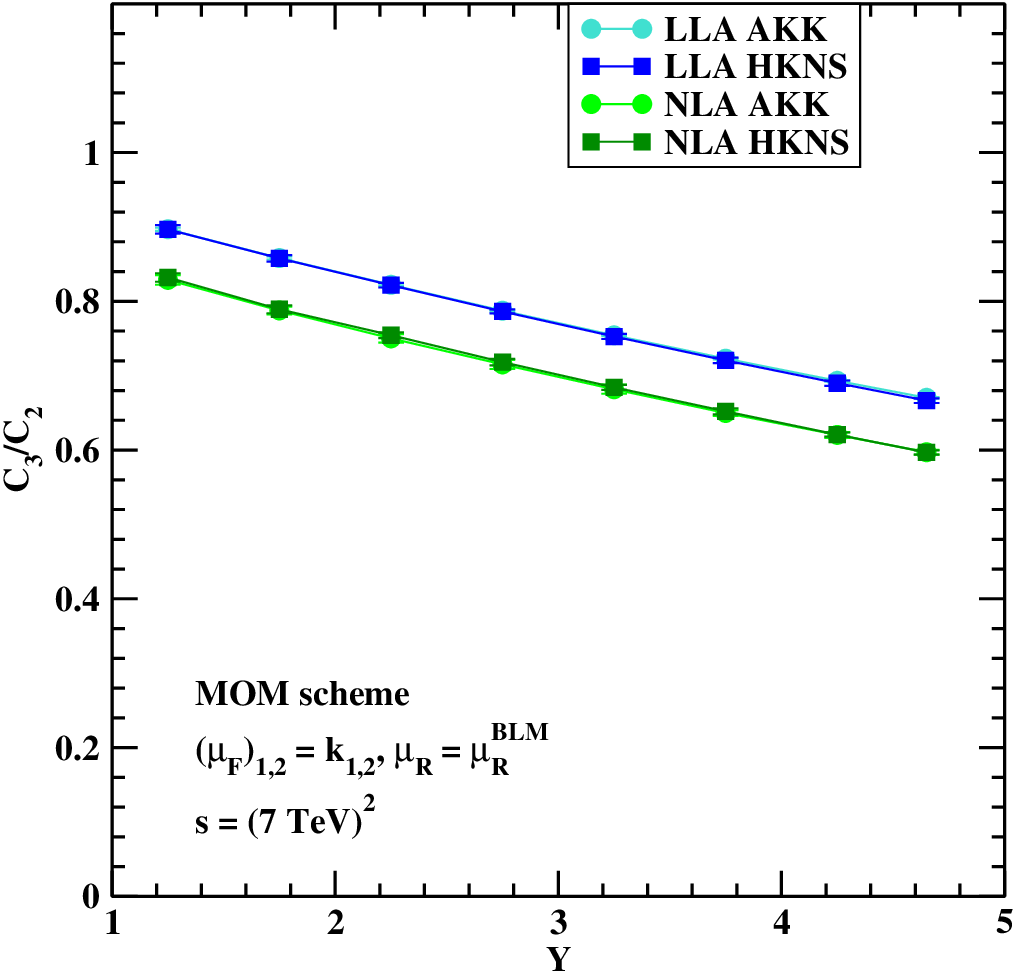}
\caption{$Y$-dependence of $C_0$ and of several ratios $C_m/C_n$ for 
$(\mu_F)_{1,2} = |\vec k_{1,2}|$, $\sqrt{s} = 7$ TeV, and $Y \leq 4.8$.}
\label{fig:ns7}
\end{figure}

For comparison, we show in Figs.~\ref{fig:ns13}~and~\ref{fig:ns7} 
the results for the same observables 
with the choice of $(\mu_F)_{1,2} = |\vec k_{1,2}|$.
The patterns we have found are very similar to the previous ones, 
but we see that the effect of having $C_1/C_0$ larger than $1$
at small $Y$ is reduced. 
Furthermore, NLA corrections are negative for larger $Y$ values.
On the basis of this, we may conclude that, in the $Y \leq 4.8$ kinematical
regime, the choice of natural scales for $\mu_F$ stabilizes the results. 

\begin{figure}[p]
\centering

   \includegraphics[scale=0.40,clip]{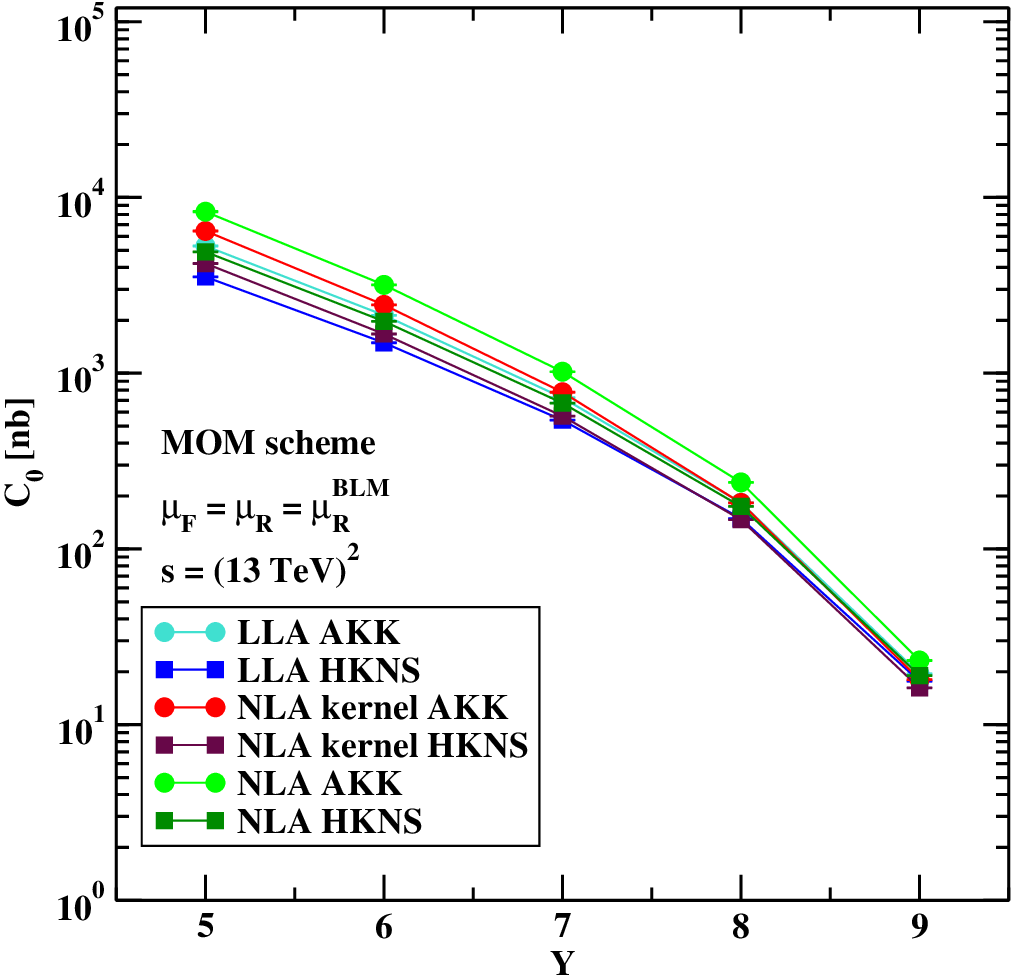}
   \includegraphics[scale=0.40,clip]{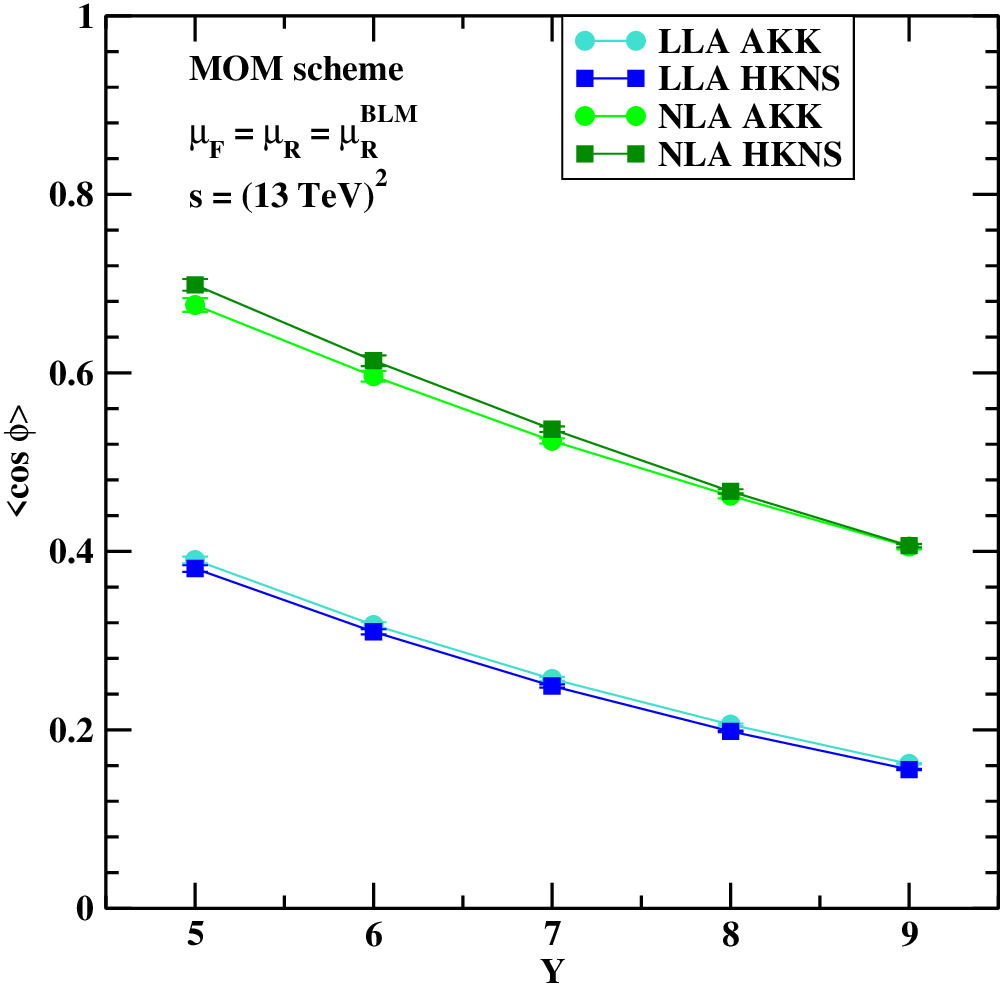}

   \includegraphics[scale=0.40,clip]{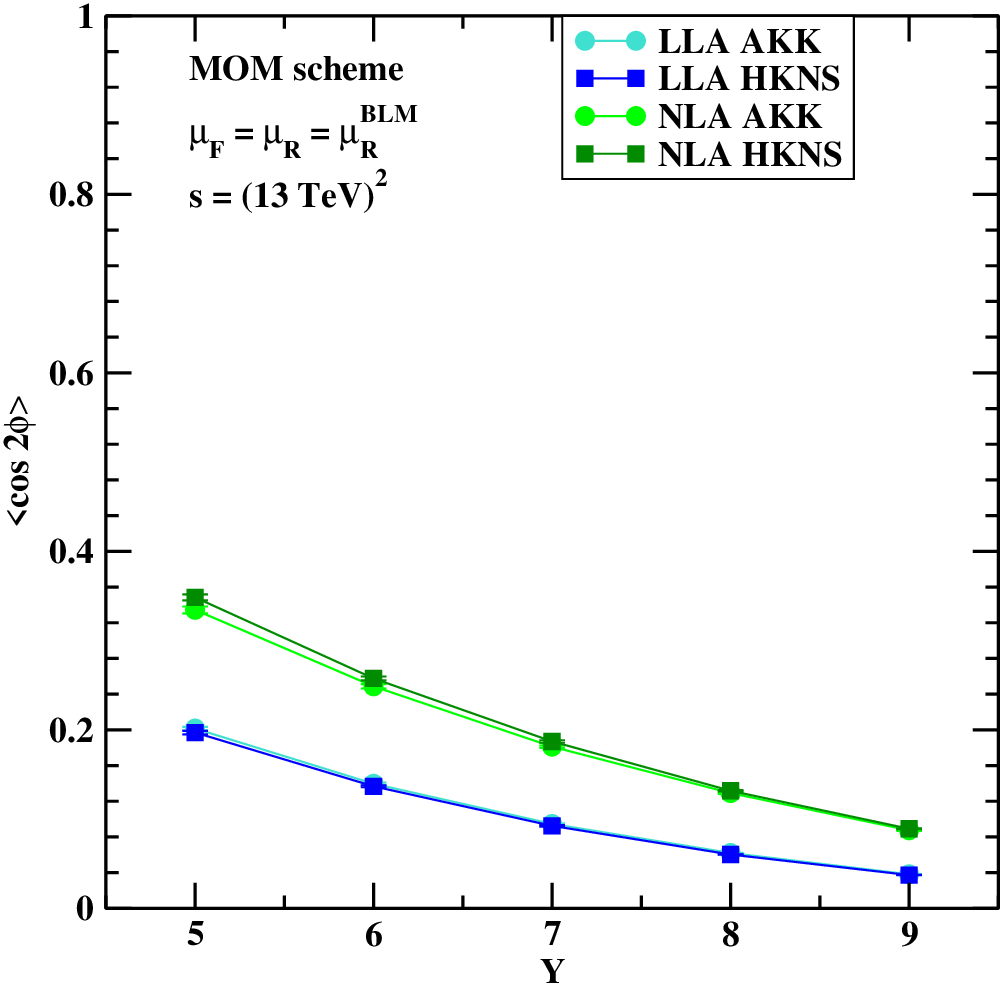}
   \includegraphics[scale=0.40,clip]{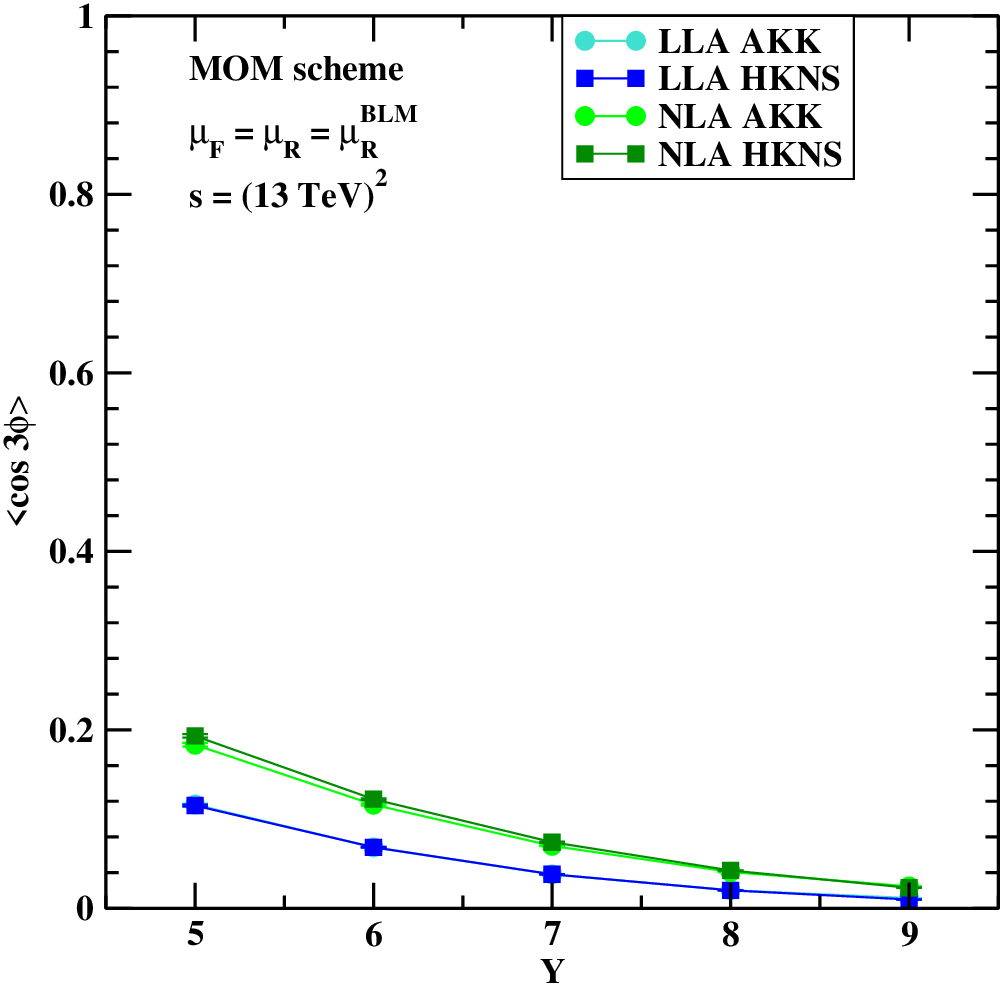}

   \includegraphics[scale=0.40,clip]{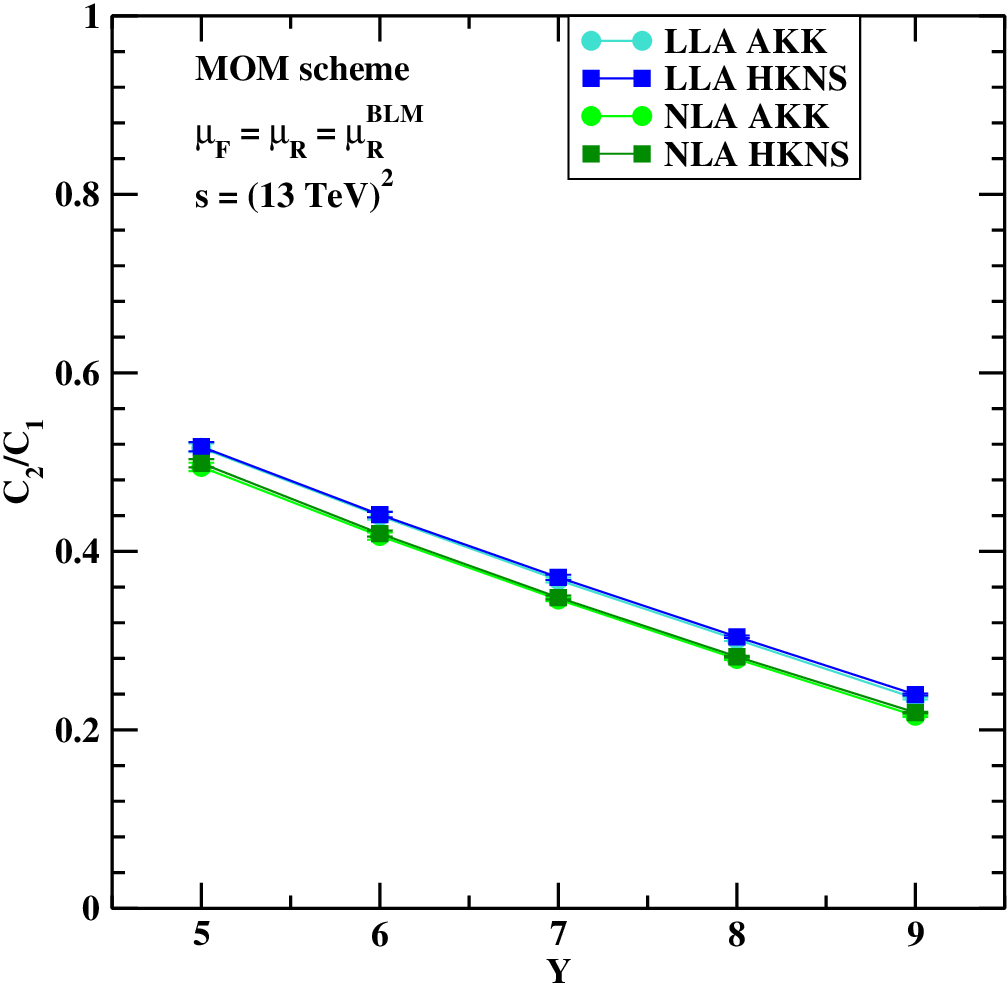}
   \includegraphics[scale=0.40,clip]{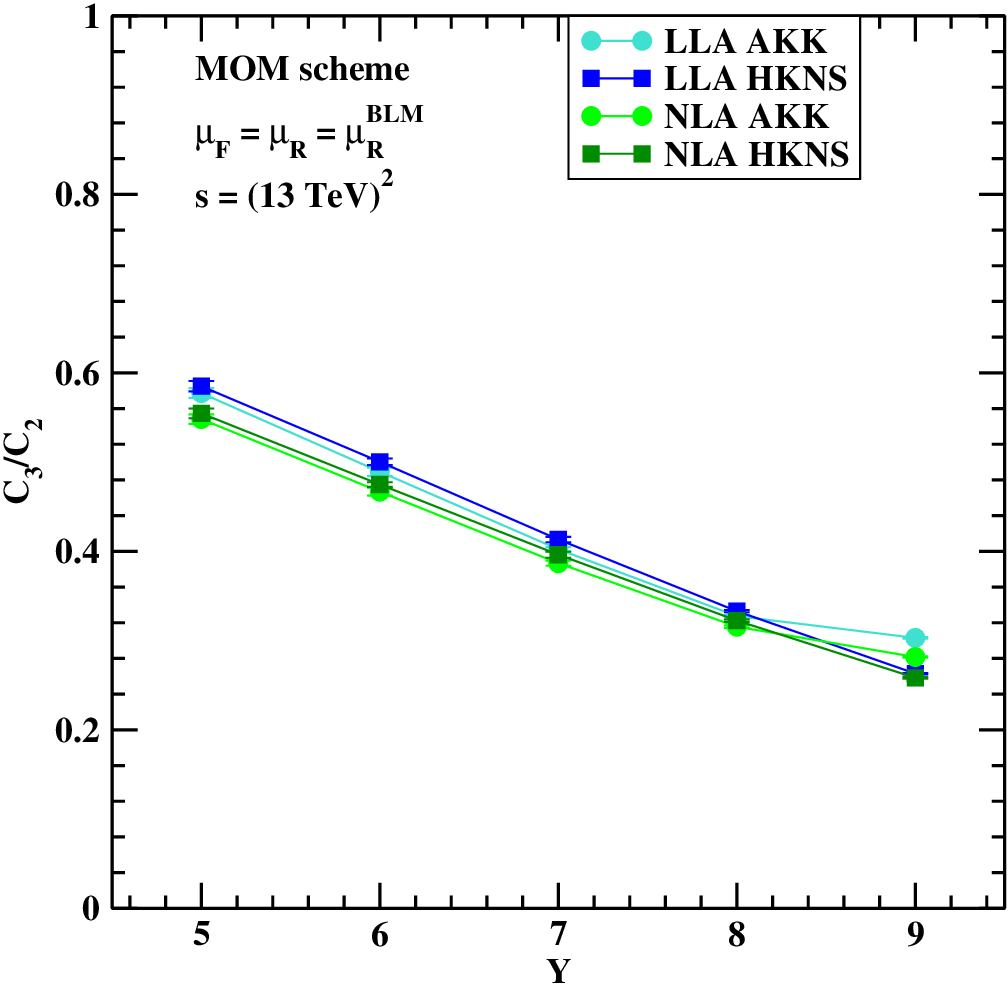}
\caption{$Y$-dependence of $C_0$ and of several ratios $C_m/C_n$ for 
$\mu_F = \mu^{\rm BLM}_R$, $\sqrt{s} = 13$ TeV, and $Y \leq 9.4$.}
\label{fig:blmLY13}
\end{figure}

\begin{figure}[p]
\centering

   \includegraphics[scale=0.40,clip]{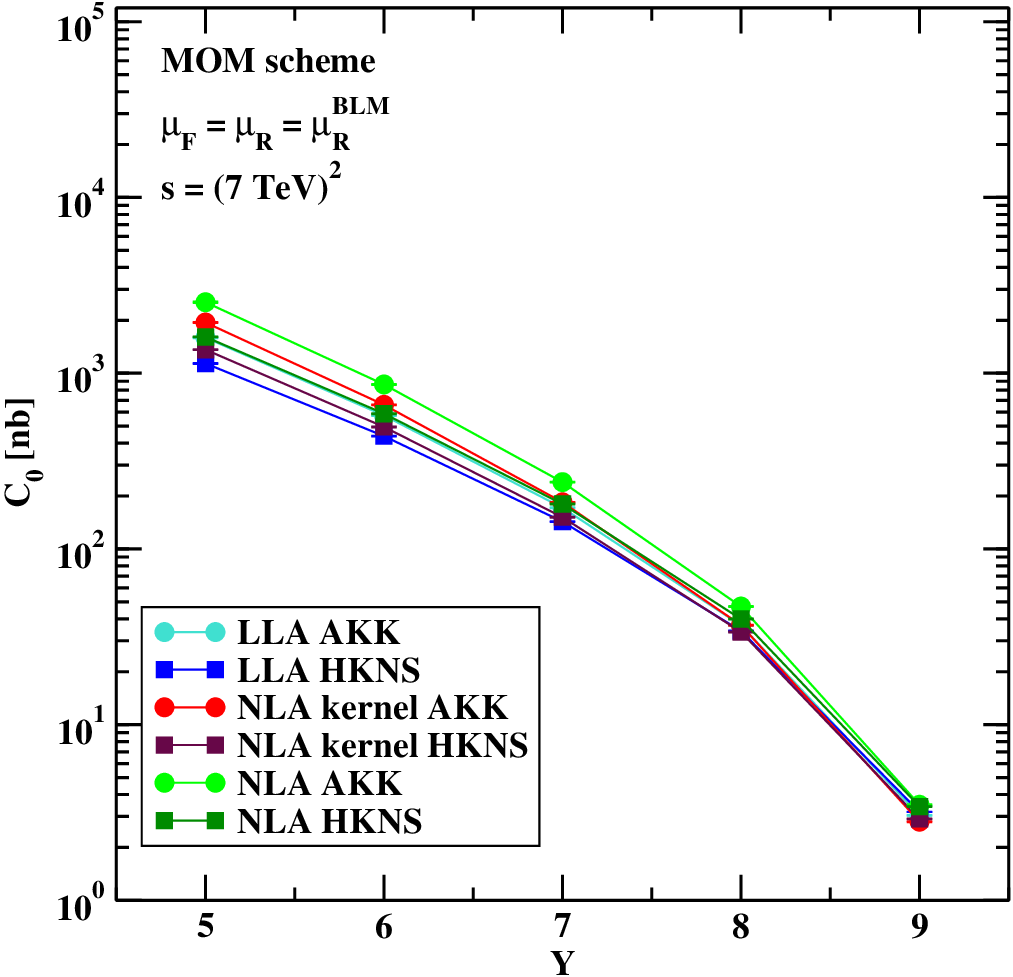}
   \includegraphics[scale=0.40,clip]{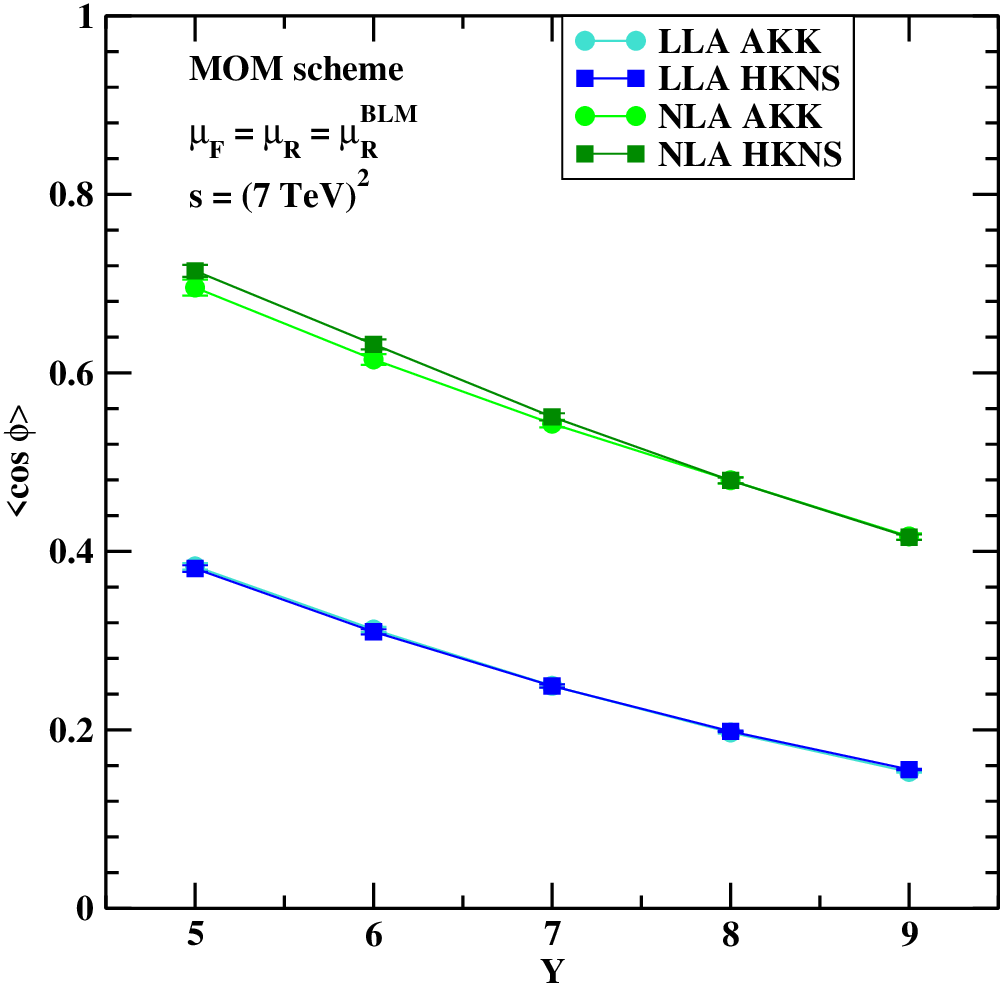}

   \includegraphics[scale=0.40,clip]{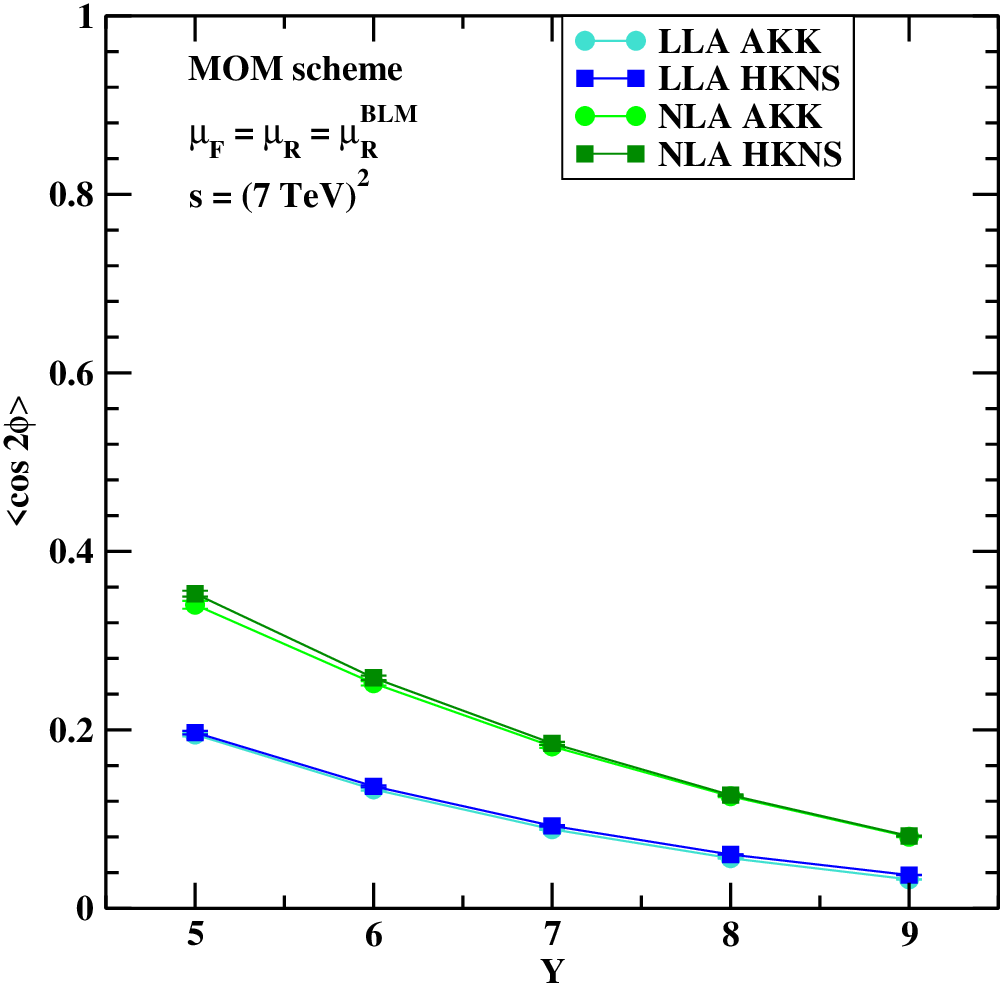}
   \includegraphics[scale=0.40,clip]{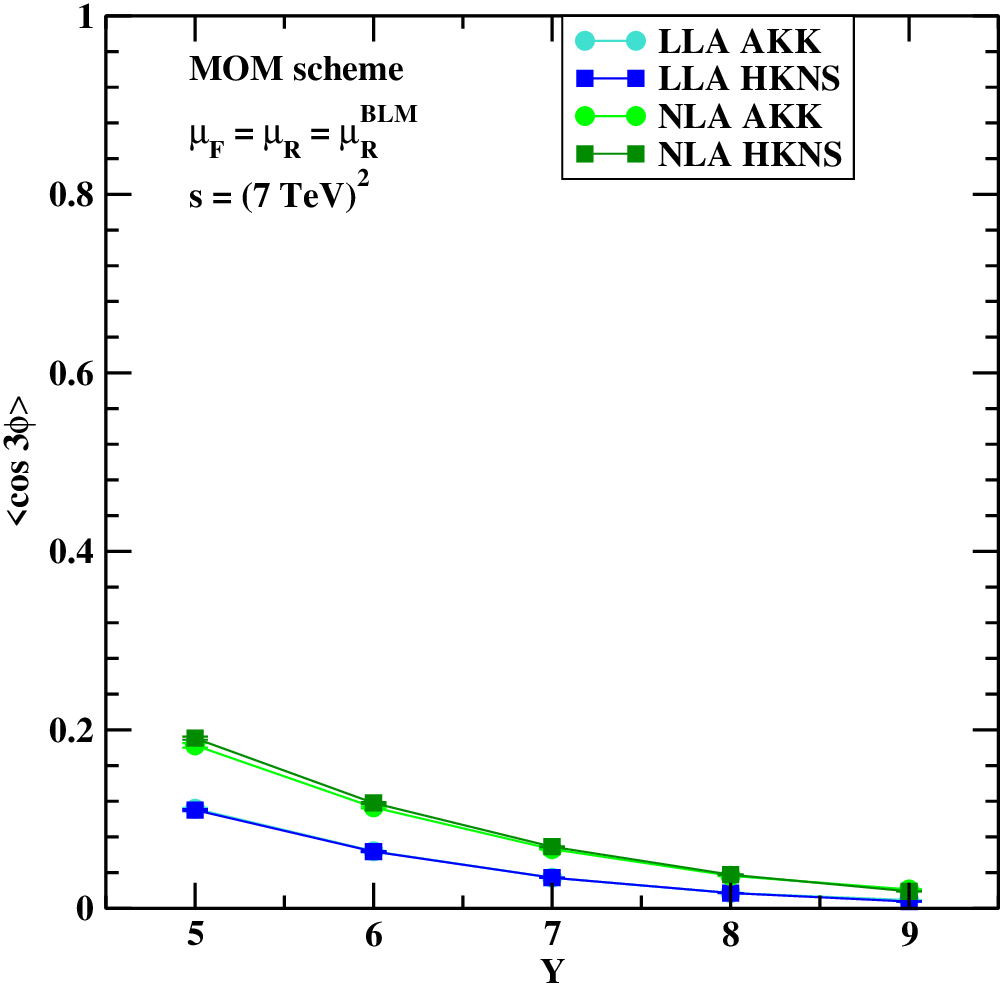}

   \includegraphics[scale=0.40,clip]{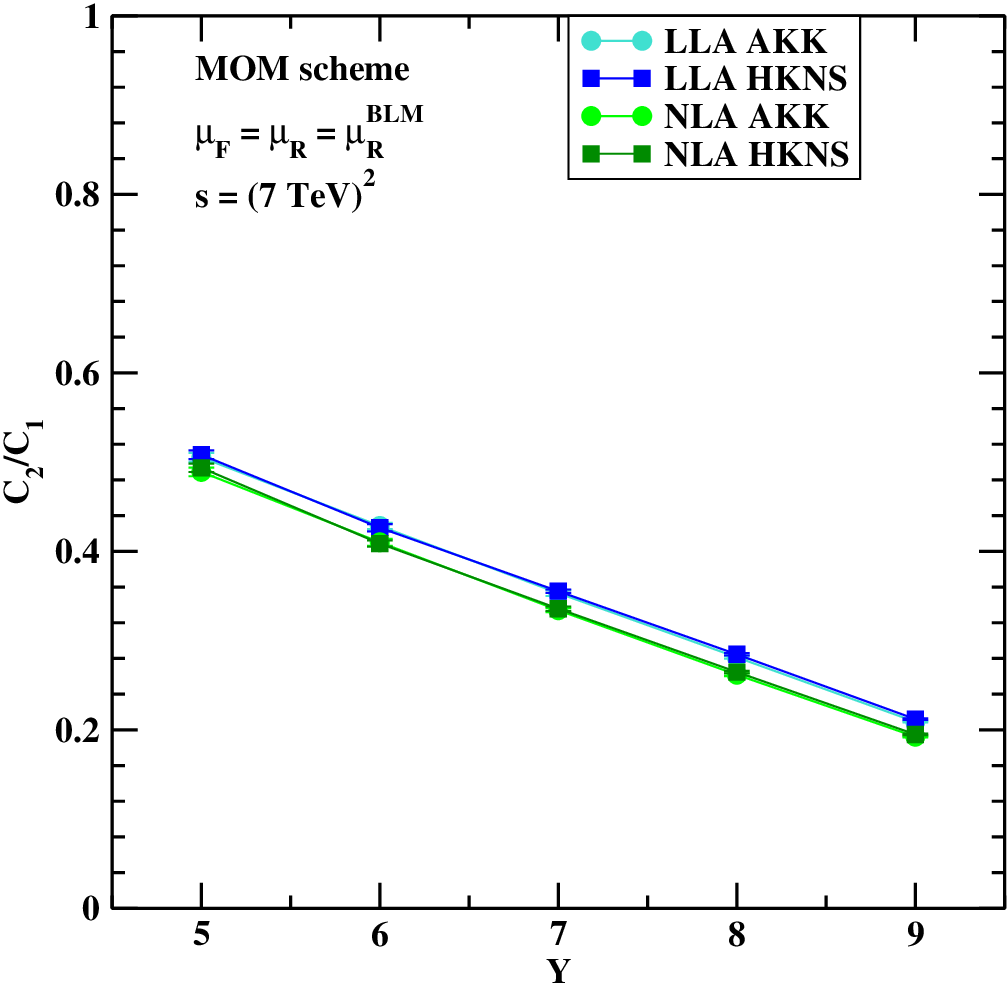}
   \includegraphics[scale=0.40,clip]{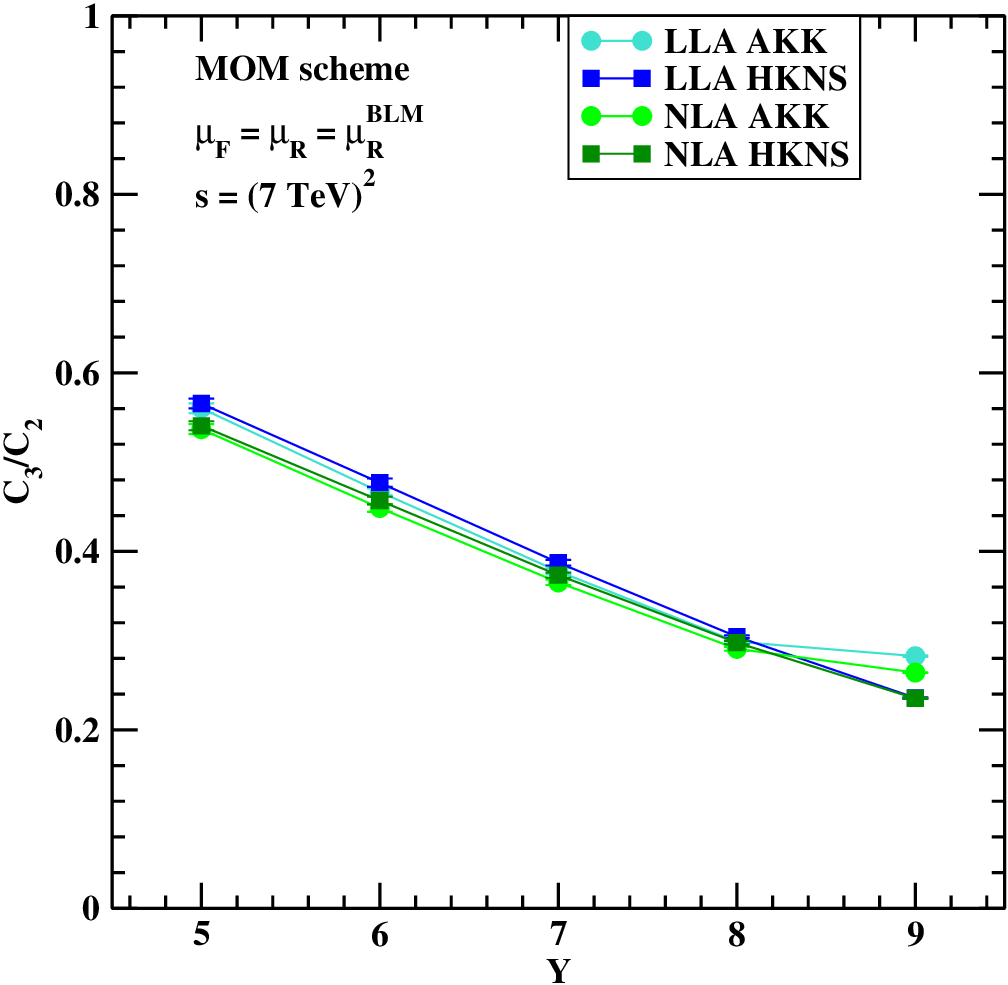}
\caption{$Y$-dependence of $C_0$ and of several ratios $C_m/C_n$ for 
$\mu_F = \mu^{\rm BLM}_R$, $\sqrt{s} = 7$ TeV, and $Y \leq 9.4$.}
\label{fig:blmLY7}
\end{figure}

In Figs.~\ref{fig:blmLY13}~and~\ref{fig:blmLY7} we present our results 
for $C_0$ and for several ratios $C_m/C_n$
at $\sqrt{s}=13$ and $7$ TeV respectively;
$\mu_F$ is set equal to $\mu^{\rm BLM}_R$,
while $Y$ lies on a larger range, {\it i.e.} $Y \leq 9.4$.

\begin{figure}[p]
\centering

   \includegraphics[scale=0.40,clip]{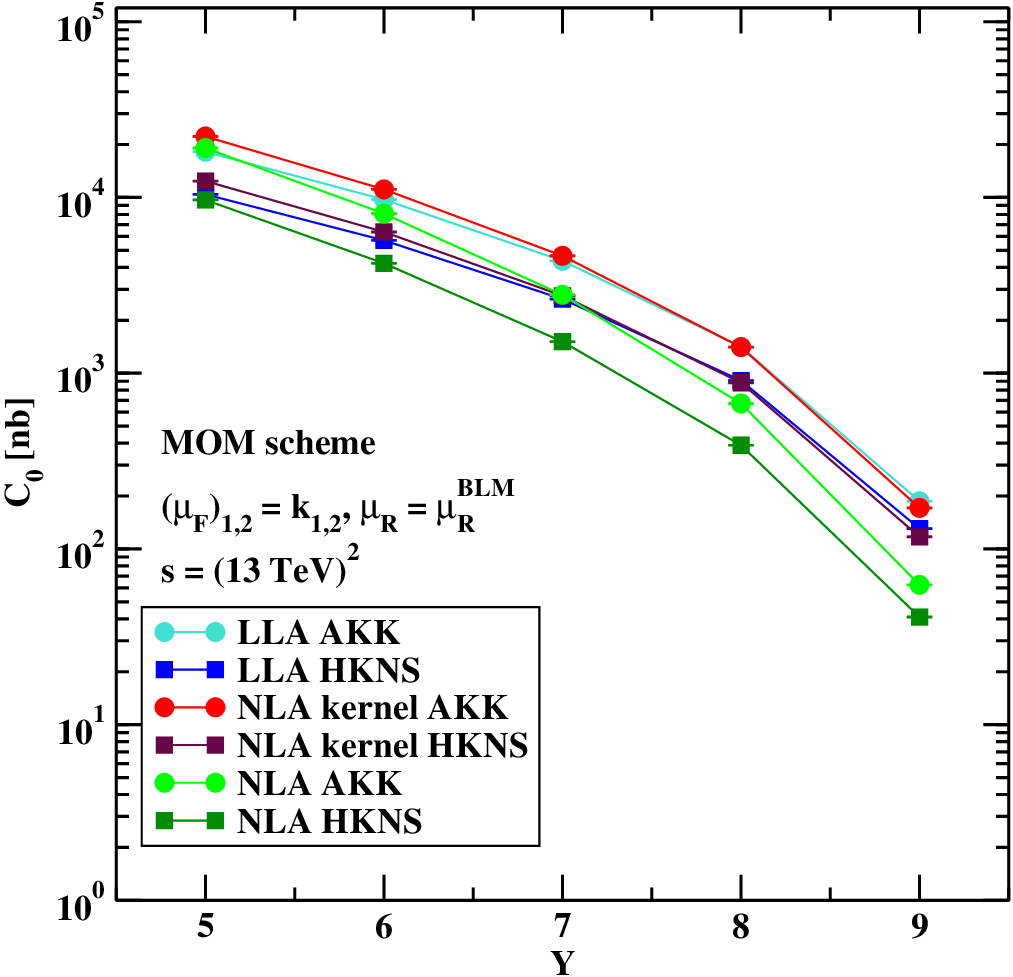}
   \includegraphics[scale=0.40,clip]{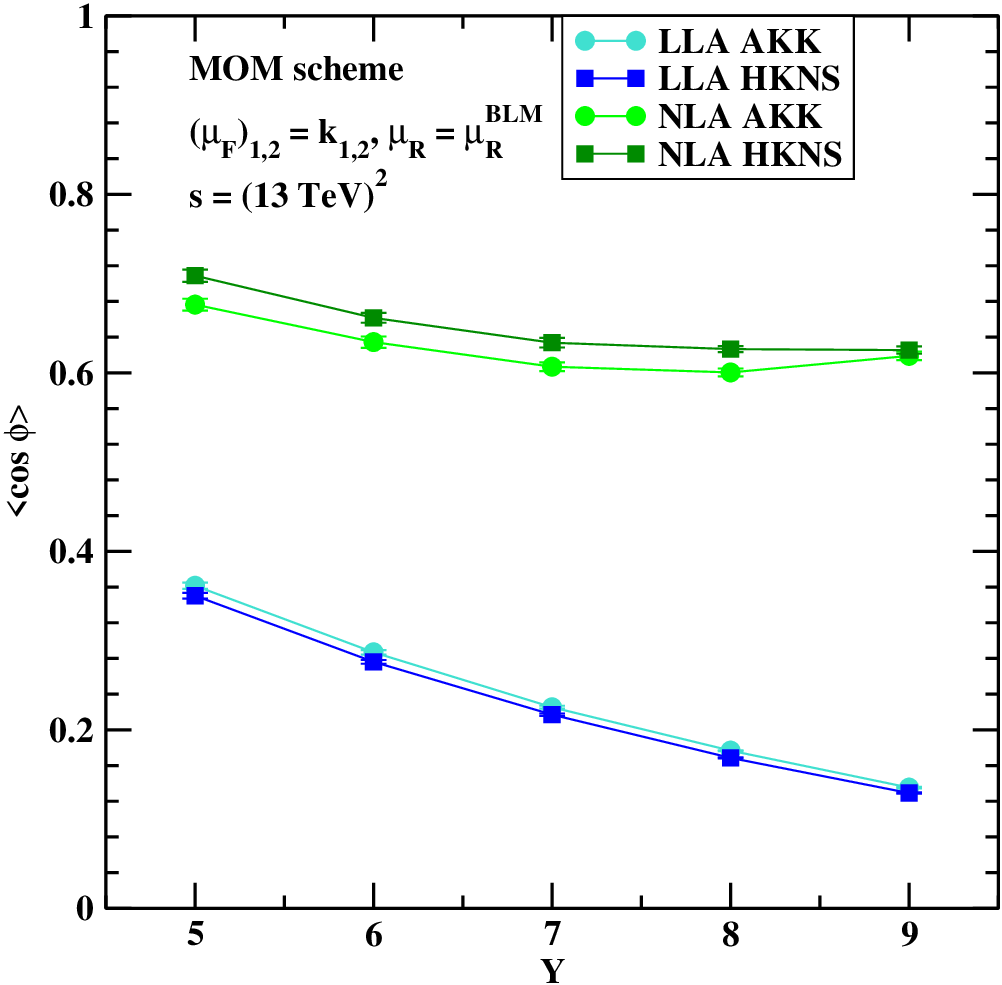}

   \includegraphics[scale=0.40,clip]{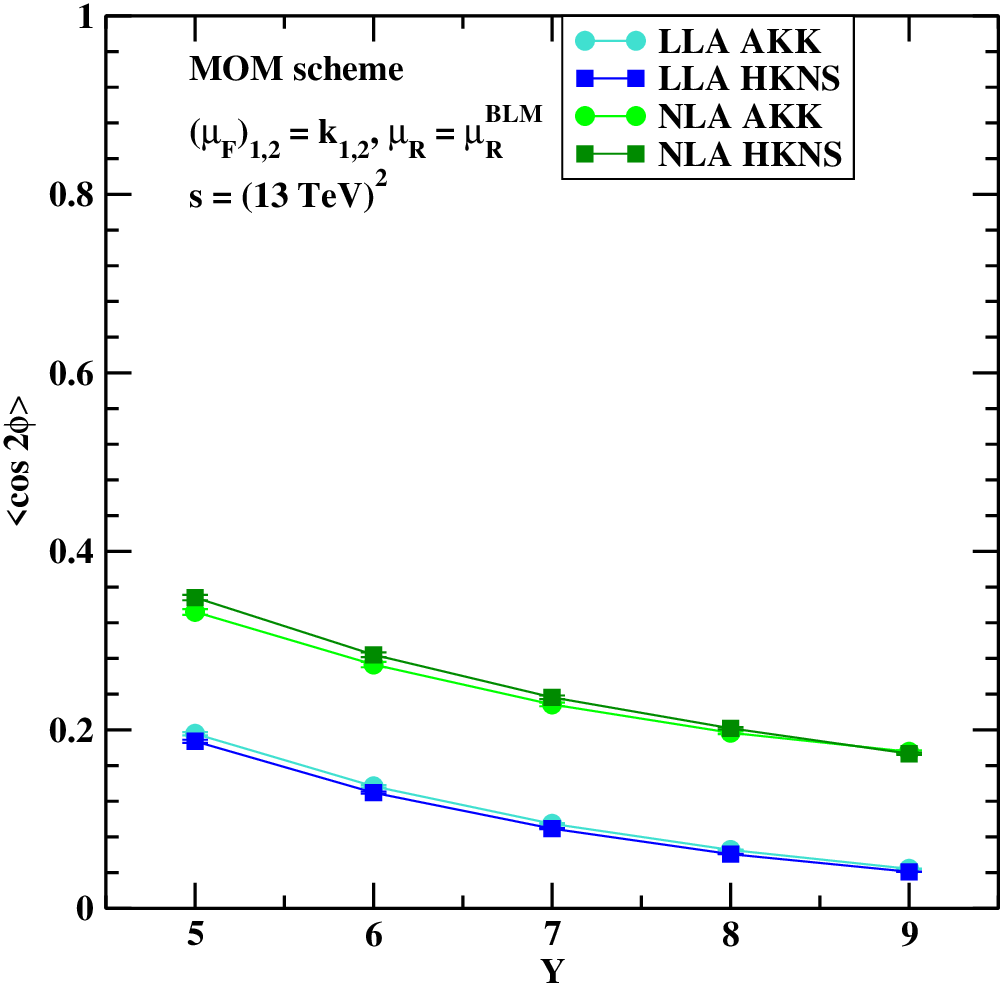}
   \includegraphics[scale=0.40,clip]{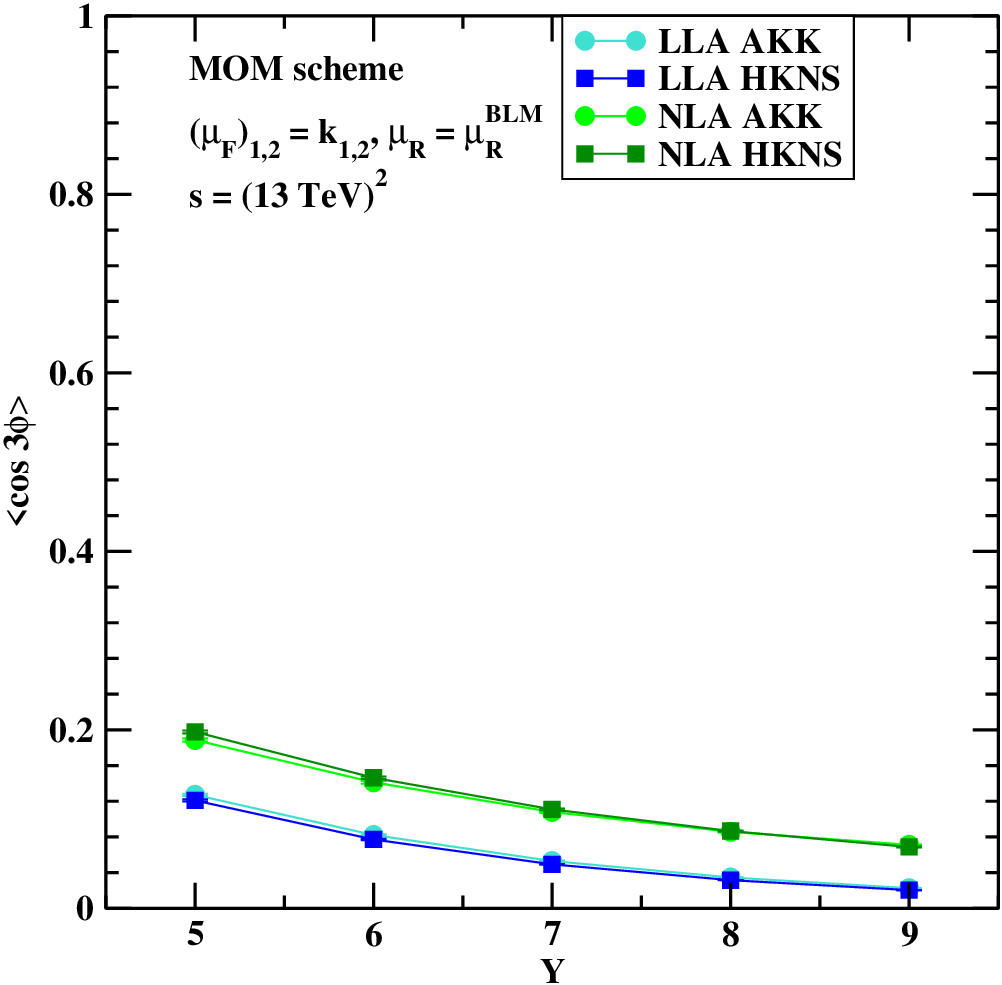}

   \includegraphics[scale=0.40,clip]{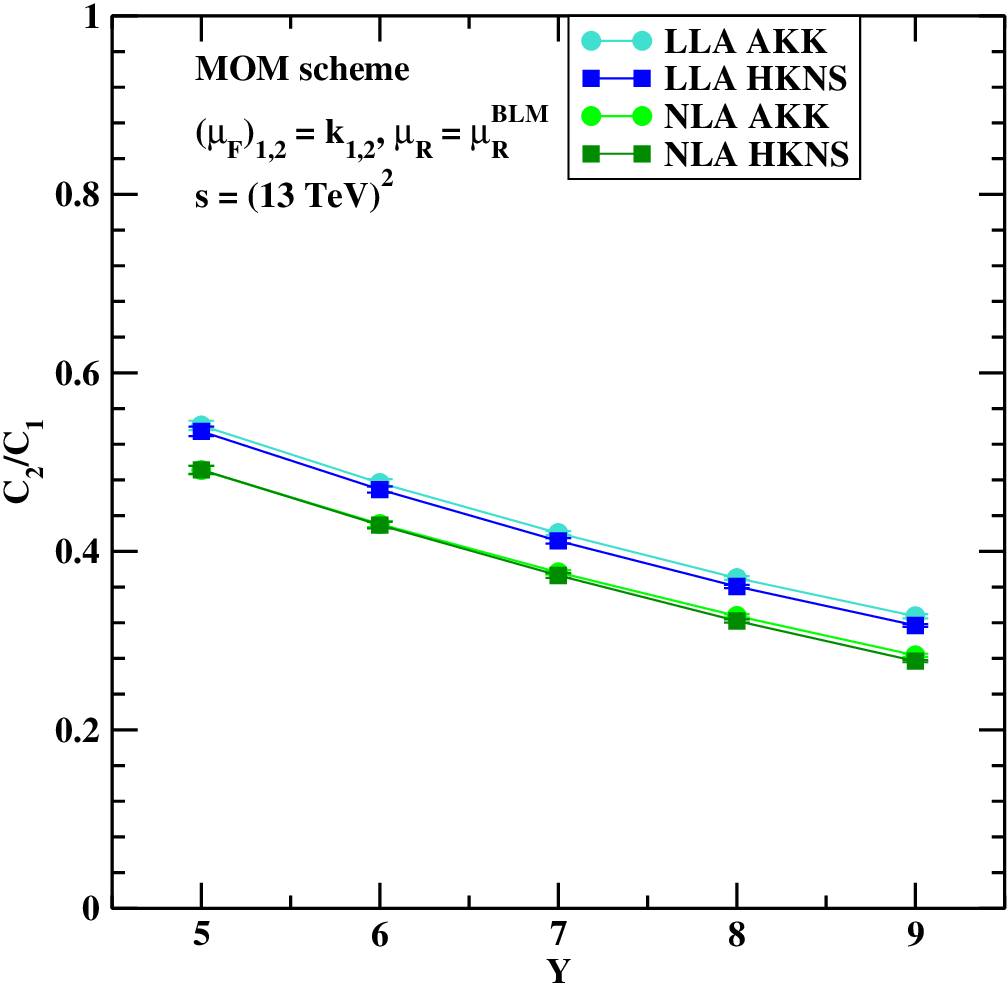}
   \includegraphics[scale=0.40,clip]{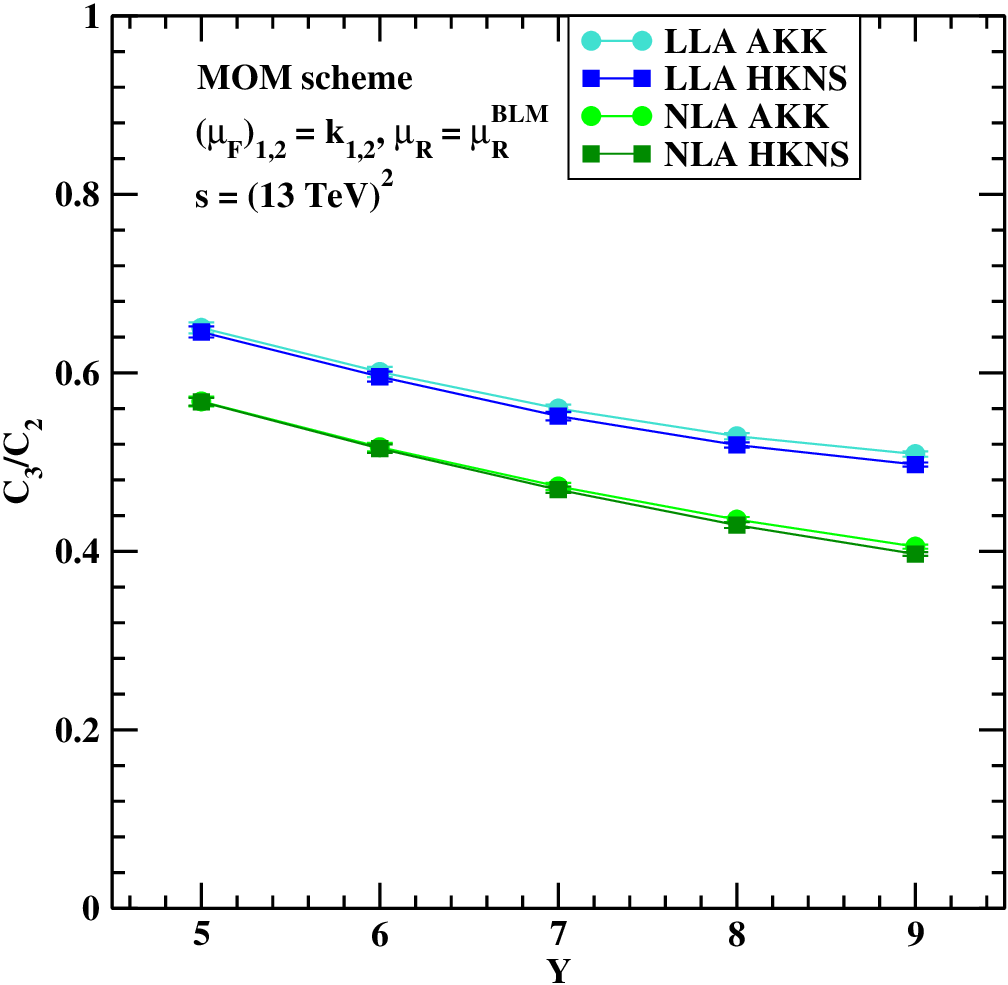}
\caption{$Y$-dependence of $C_0$ and of several ratios $C_m/C_n$ for 
$(\mu_F)_{1,2} = |\vec k_{1,2}|$, $\sqrt{s} = 13$ TeV, and $Y \leq 9.4$.}
\label{fig:nsLY13}
\end{figure}

For comparison, we show in Figs.~\ref{fig:nsLY13}~and~\ref{fig:nsLY7} 
the results for the same observables 
with the choice of $(\mu_F)_{1,2} = |\vec k_{1,2}|$.
We clearly see that, in the case of larger rapidity intervals $Y$ and with the
natural choice for the factorization scale, the situation is different in
comparison to the $\mu_F=\mu_R^{BLM}$ choice: the NLA corrections to the cross
section $C_0$ are negative, while the pattern of $C_1/C_0$ shows a somewhat
unexpected ``turn-up'' at large $Y$, and these effects are more pronounced for
the lower LHC energy,  $\sqrt{s}=7 \ \rm{TeV}$.
Such a sensitivity to the factorization scale setting may be an indication of
the fact that with the increase of $Y$ values we are moving towards the
threshold region, where the energy of detected dihadron system becomes
comparable with  $\sqrt{s}$. In this situation the FFs and PDFs are probed
in regions that are close to the end-points of their definitions, where they
exhibit large dependence on the factorization scale. From the physical site,  
in this kinematics the undetected hard-gluon radiation is getting restricted
and only radiation of soft gluons is allowed.
Soft-gluon radiation can not change the kinematics of the hard subprocess,
therefore one expects restoration of the correlation of the detected dihadrons
in the relative azimuthal angle when we  approach  the threshold region.    
It is well known that in this situation large threshold double logarithms
appear in the perturbative series, and such contributions have to be resummed
to all orders. Resummation in the kinematics where both threshold and BFKL
logarithms are important is an interesting task, but it goes well beyond
the scope of the present study. Here we just note that pure BFKL
predictions in the region of largest $Y$ become rather sensitive to the
choice of the factorization scale. 

To better assess the factorization scale dependence, we have considered
also the case when $\mu_F$ is varied around its ``natural value''
$\sqrt{\vec k_1 \vec k_2}$ by a factor $r$ taking values in the range 1/2 to
four. In Fig.~\ref{fig:muf}, as a selection of our results, we present the plots
for $C_0$ and $C_1/C_0$ at a squared center-of-mass energy of 7 and 13~TeV
for the rapidity region $Y\leq 4.8$ and the HKNS parametrization of the
fragmentation functions.

At the end of this section it is worth to note that the general features
of our predictions for dihadron production are rather similar to those
obtained earlier for the Mueller-Navelet jet process. Although the BFKL 
resummation leads to the growth with energy of the partonic subprocess
cross sections, the convolution of the latter with the proton PDFs makes
the net effect of a decrease with $Y$ of our predictions. This is due to 
the fact that, at larger values of $Y$, PDFs are probed effectively at
larger values of $x$, where they fall very fast.
For the dihadron azimuthal correlations we predict a decreasing behavior
with $Y$. That originates from the increasing amount of hard undetected
parton radiation in the final state allowed by the growth of the partonic 
subprocess energy. 

\section{Conclusions and Outlook}

In this paper we studied the inclusive dihadron production
process at the LHC within the BFKL approach,   
giving the first complete phenomenological predictions 
for cross sections and azimuthal correlation momenta 
in the full NLA approximation.
We implemented the exact version of the BLM optimization procedure, which 
requires the choice of renormalization scale $\mu_R=\mu_R^{BLM}$ such that   
it makes completely vanish the NLA terms proportional to the QCD
$\beta$-function.\footnote{To avoid misunderstandings, by ``exact
  implementation of the BLM procedure'', we mean here that with our choice of
  the renormalization scale all terms proportional to the QCD $\beta_0$ vanish
  within the accuracy of our calculation, NLA BFKL resummation. To get
  such full cancellation, the terms originated both from the NLA kernel
  and the hadron vertices have to be taken into account -- see discussion
  after Eq.~(\ref{beta0}). } 
This procedure leads to rather large values of the scale $\mu_R^{BLM}$ and
it allows to minimize the size of the NLA corrections in our observables.
We considered two center-of-mass energies, $\sqrt s = 7, 13$ TeV, 
and two different ranges for the rapidity interval between the two hadrons
in the final state, $Y \leq 4.8$ and $Y \leq 9.4$, 
which are typical for the last CMS analyses.
The first rapidity range we investigated, $Y\leq 4.8$, may look to 
be not large enough for the dominance of BFKL dynamics. But we see, however, 
that in this range there are large NLA BFKL corrections, thus indicating that
the BFKL resummation is playing here a non-trivial role. To clarify the issue  
it would be very interesting to confront our predictions with the results of 
fixed-order NLO DGLAP calculations. But this would require new numerical 
analysis in our semihard kinematic range, because the existing NLO DGLAP 
results cover the hard kinematic range for the energies of fixed target 
experiments, see for instance~\cite{Owens:2001rr,Almeida:2009jt}.

As for the hadron's transverse momenta, we imposed the symmetrical lower
cutoff: $|\vec k_{1,2}|\geq 5$~GeV. 
Considering a region of lower hadron transverse momenta, 
say $|\vec k_{1,2}| \geq 2$~GeV, would lead to even larger values of the cross
sections. 
But it should be noted that in our calculation we use the BFKL method together 
with leading-twist collinear factorization, which means that we 
are systematically neglecting power-suppressed corrections. 
Therefore, going to smaller transverse momenta we would enter a region 
where higher-twist effects must be important. 

The general features of our predictions for dihadron production are
rather similar to those obtained earlier for the Mueller-Navelet jet process.
In particular, we observe that the account of NLA BFKL terms leads to much
less azimuthal angle decorrelation with increasing $Y$ in comparison to
LLA BFKL calculations.
As for the difference between the Mueller-Navelet jet and dihadron
production processes, we would mention the fact that, contrary to the jets'
case, the full account of NLA terms leads in dihadron production to an
increase of our predictions for the cross sections in comparison to the
LLA BFKL calculation.

We considered the effect of using different parametrization sets 
for the PDFs and the FFs, that could potentially give rise 
to uncertainties which, in principle, are not negligible. 
We did some preliminary tests devoted to gauge 
the effect of using different PDF routines, 
showing that it leads to no significant difference in the results.
Then, we investigated the $Y$-behavior of our observables 
by using two different FF parametrizations. 
Our calculation with the AKK FFs gives bigger cross sections, while the
difference between AKK and HKNS is small, since the FFs uncertainties are mostly
wiped out in the azimuthal ratios.

We studied the effect of using two different choices 
for the factorization scale, $\mu_F = \mu_R^{\rm BLM}$ and
$(\mu_F)_{1,2} = |\vec k_{1,2}|$, whereas $\mu_R = \mu_R^{\rm BLM}$
runs at BLM scales.
We see  some  difference in predictions within these two approaches,
especially for larger values of $Y$ and at the smaller value of the energy
 $\sqrt{s}=7 \ \rm{TeV}$. In this region, the kinematic restriction for
the undetected hard gluon radiation may start to be important, requiring
resummation of threshold double logs together with BFKL logarithms of energy.
This issue maybe a physical reason for the observed strong dependence on the
factorization scale choice in our pure BFKL approach, and it definitely
deserves a further study.

The applicability border for our approach could be established either
by comparing our predictions with future data or by confronting it with some
other theoretical predictions which do include higher-twist effects. 
For the last point, one can consider an alternative, higher-twist production 
mechanism, related with multiparton interactions in QCD (for a review, 
see~\cite{Diehl:2011yj}). The double-parton scattering contribution to the  
Mueller-Navelet jet production was considered in the 
papers~\cite{Ducloue:2015jba} and~\cite{Maciula:2014pla}, using different 
approaches. It would be very interesting if similar estimates were done 
also for the case of dihadron production. 

We plan to extend this study by investigating the effect 
of using asymmetrical cuts for the hadrons' transverse momenta 
as well as studying less inclusive processes where at least 
one light charged hadron is always tagged in the final state.

We encourage experimental collaborations to include the
study of the dihadron production in the program of future analyses 
at the LHC, making use of a new suitable channel to improve our knowledge
about the dynamics of strong interactions in the Regge limit.

\section{Acknowledgments}

We thank G.~Safronov and I.~Khmelevskoi for stimulating and helpful discussions.
This work was supported in part by the RFBR-15-02-05868.

\begin{figure}[p]
\centering

   \includegraphics[scale=0.40,clip]{C0_ns_LY_7TeV.eps}
   \includegraphics[scale=0.40,clip]{C1C0_ns_LY_7TeV.eps}

   \includegraphics[scale=0.40,clip]{C2C0_ns_LY_7TeV.eps}
   \includegraphics[scale=0.40,clip]{C3C0_ns_LY_7TeV.eps}

   \includegraphics[scale=0.40,clip]{C2C1_ns_LY_7TeV.eps}
   \includegraphics[scale=0.40,clip]{C3C2_ns_LY_7TeV.eps}
\caption{$Y$-dependence of $C_0$ and of several ratios $C_m/C_n$ for 
$(\mu_F)_{1,2} = |\vec k_{1,2}|$, $\sqrt{s} = 7$ TeV, and $Y \leq 9.4$.}
\label{fig:nsLY7}
\end{figure}

\begin{figure}[p]
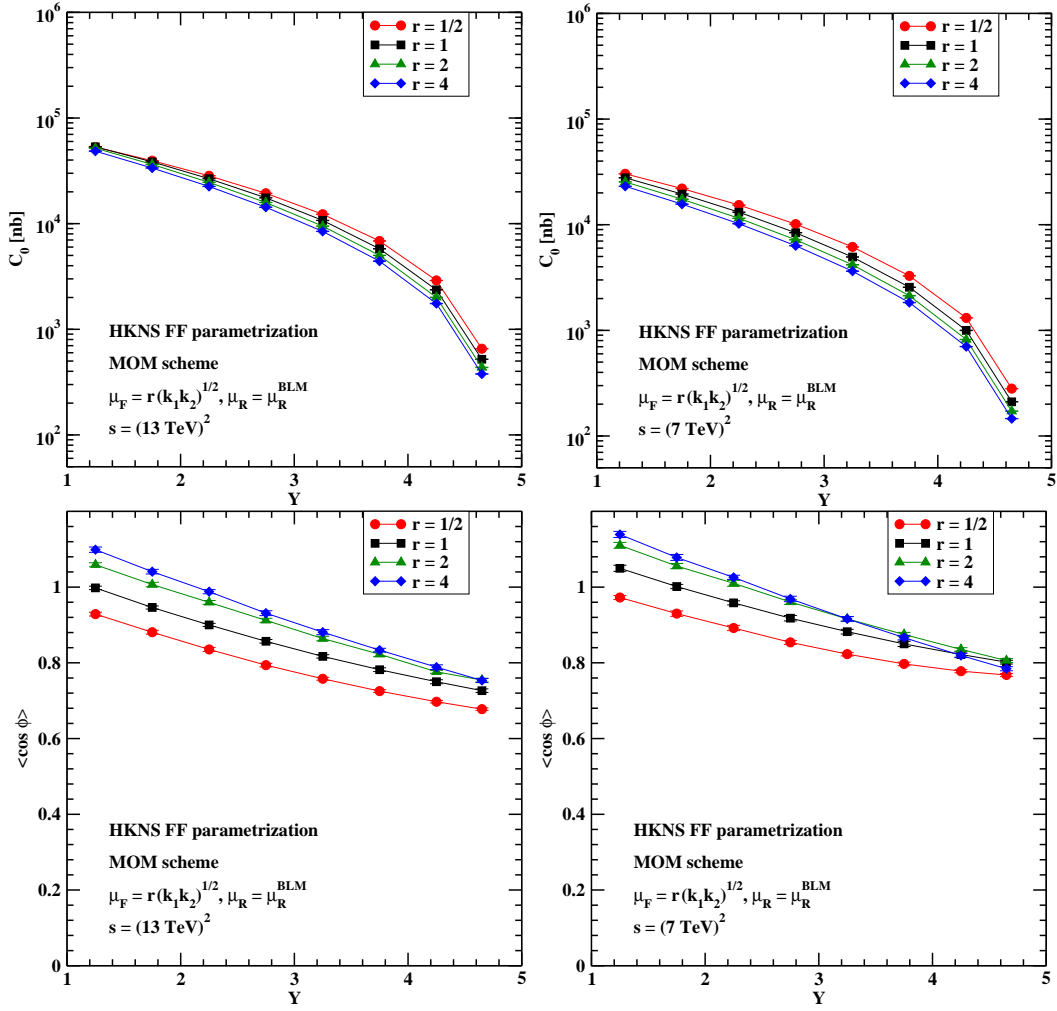

\centering

   \includegraphics[scale=0.40,clip]{C0_muf_13TeV.eps}
   \includegraphics[scale=0.40,clip]{C0_muf_7TeV.eps}

   \includegraphics[scale=0.40,clip]{C1C0_muf_13TeV.eps}
   \includegraphics[scale=0.40,clip]{C1C0_muf_7TeV.eps}
\caption{$Y$-dependence of $C_0$ and of $C_1/C_0$ for 
  $(\mu_F)_{1,2} = r \sqrt{\vec k_1 \vec k_2} $, with $r$ =1/2, 1, 2, 4, and
  $Y \leq 4.8$.}
\label{fig:muf}
\end{figure}


\begin{thebibliography}{99}

\bibitem{BFKL}
V.S.~Fadin, E.~Kuraev, L.~Lipatov, Phys. Lett. B \textbf{60}, 50 (1975);
%
Sov. Phys. JETP \textbf{44}, 443 (1976);
%
E.~Kuraev, L.~Lipatov, V.S.~Fadin, Sov. Phys. JETP \textbf{45}, 199 (1977);
%
I.~Balitsky, L.~Lipatov, Sov. J. Nucl. Phys. \textbf{28}, 822 (1978).

\bibitem{Fadin:1998py}
  V.S.~Fadin, L.N.~Lipatov,
  Phys.\ Lett.\ B {\bf 429} (1998) 127
  [hep-ph/9802290].

\bibitem{Ciafaloni:1998gs}
  M.~Ciafaloni, G.~Camici,
  Phys.\ Lett.\ B {\bf 430} (1998) 349
  [hep-ph/9803389].

\bibitem{Fadin:1998jv}
  V.S.~Fadin, R.~Fiore, A.~Papa,
  Phys.\ Rev.\ D {\bf 60} (1999) 074025
  [hep-ph/9812456].

\bibitem{FG00}
V.S. Fadin, D.A. Gorbachev, Pisma v Zh. Eksp. Teor. Fiz. {\bf 71} (2000) 322
[JETP Letters {\bf 71} (2000) 222]; Phys. Atom. Nucl. {\bf 63} (2000) 2157
[Yad. Fiz. {\bf 63} (2000) 2253].

\bibitem{FF05}
V.S.~Fadin, R.~Fiore, Phys. Lett. {\bf B610} (2005) 61
[{\it Erratum-ibid.} {\bf 621} (2005) 61] [hep-ph/0412386];
Phys. Rev. D {\bf 72} (2005) 014018 [hep-ph/0502045].

\bibitem{Hentschinski:2012kr}
  M.~Hentschinski, A.~Sabio Vera, C.~Salas,
  [arXiv:1209.1353 [hep-ph]].

\bibitem{Hentschinski:2013id}
  M.~Hentschinski, A.~Sabio Vera, C.~Salas,
  Phys.\ Rev.\ D {\bf 87} (2013) 076005
  
\bibitem{Mueller:1986ey}
A.H.~Mueller, H.~Navelet, 
Nucl. Phys. B \textbf{282}, 727 (1987).

\bibitem{Bartels:2001ge}
  J.~Bartels, D.~Colferai, G.P.~Vacca,
  Eur.\ Phys.\ J.\ C {\bf 24} (2002) 83
  [hep-ph/0112283].

\bibitem{Bartels:2002yj}
  J.~Bartels, D.~Colferai, G.P.~Vacca,
  Eur.\ Phys.\ J.\ C {\bf 29} (2003) 235
  [hep-ph/0206290].

\bibitem{Caporale:2011cc}
F.~Caporale, D.Yu.~Ivanov, B.~Murdaca, A.~Papa and A.~Perri,
JHEP {\textbf 1202}, 101 (2012). 

\bibitem{Ivanov:2012ms}
D.Yu.~Ivanov, A.~Papa,
JHEP {\bf 1205}, 086 (2012).

\bibitem{Colferai:2015zfa}
D.~Colferai, A.~Niccoli,
JHEP {\bf 1504}, 071 (2015).

\bibitem{DGLAP}
V.N.~Gribov, L.N.~Lipatov, Sov. J. Nucl. Phys. {\bf 15} (1972) 438;
G.~Altarelli, G.~Parisi, Nucl. Phys. B {\bf 126} (1977) 298;
Y.L.~Dokshitzer, Sov. Phys. JETP {\bf 46} (1977) 641.

\bibitem{Colferai:2010wu}
  D.~Colferai, F.~Schwennsen, L.~Szymanowski, S.~Wallon,
  JHEP {\bf 1012} (2010) 026
  [arXiv:1002.1365 [hep-ph]].

\bibitem{Angioni:2011wj} 
  M.~Angioni, G.~Chachamis, J.D.~Madrigal, A.~Sabio~Vera,
  Phys.\ Rev.\ Lett.\  {\bf 107}, 191601 (2011)
  [arXiv:1106.6172 [hep-th]].

\bibitem{Ducloue:2013wmi}
  B.~Duclou\'e, L.~Szymanowski, S.~Wallon,
  JHEP {\bf 1305} (2013) 096
  [arXiv:1302.7012 [hep-ph]].

\bibitem{Caporale:2012ih}
  F.~Caporale, D.Yu.~Ivanov, B.~Murdaca, A.~Papa,
  Nucl.\ Phys.\ B {\bf 877} (2013) 73
  [arXiv:1211.7225 [hep-ph]].

\bibitem{Caporale:2013uva}
  F.~Caporale, B.~Murdaca, A.~Sabio Vera, C.~Salas,
  Nucl.\ Phys.\ B {\bf 875} (2013) 134
  [arXiv:1305.4620 [hep-ph]].

\bibitem{Ducloue:2013bva}
  B.~Duclou\'e, L.~Szymanowski, S.~Wallon,
  Phys.\ Rev.\ Lett.\  {\bf 112} (2014) 082003
  [arXiv:1309.3229 [hep-ph]].

\bibitem{Ducloue:2014koa}
  B.~Duclou\'e, L.~Szymanowski, S.~Wallon,
  Phys.\ Lett.\ B {\bf 738} (2014) 311
  [arXiv:1407.6593 [hep-ph]].

\bibitem{Caporale:2014gpa}
F.~Caporale, D.Yu.~Ivanov, B.~Murdaca, A.~Papa,
Eur.\ Phys.\ J.\ C {\bf 74}, no. 10, 3084 (2014)
[Eur.\ Phys.\ J.\ C {\bf 75}, no. 11, 535 (2015)].

\bibitem{Ducloue:2015jba}
  B.~Duclou\'e, L.~Szymanowski and S.~Wallon,
  Phys.\ Rev.\ D {\bf 92} (2015) no.7,  076002
  [arXiv:1507.04735 [hep-ph]].

\bibitem{Caporale:2015uva}
  F.~Caporale, D.Yu.~Ivanov, B.~Murdaca, A.~Papa,
  Phys.\ Rev.\ D {\bf 91} (2015) no.11,  114009
  [arXiv:1504.06471 [hep-ph]].

\bibitem{Celiberto:2015yba}
  F.G.~Celiberto, D.Yu.~Ivanov, B.~Murdaca, A.~Papa,
  Eur.\ Phys.\ J.\ C {\bf 75} (2015) no.6,  292
  [arXiv:1504.08233 [hep-ph]].

\bibitem{Celiberto:2016ygs}
  F.G.~Celiberto, D.Yu.~Ivanov, B.~Murdaca, A.~Papa,
  Eur.\ Phys.\ J.\ C {\bf 76} (2016) no.4,  224
  [arXiv:1601.07847 [hep-ph]].

\bibitem{Chachamis:2015crx}
  G.~Chachamis,
  arXiv:1512.04430 [hep-ph].
  
\bibitem{DelDuca:1993mn}
V.~Del~Duca, C.R.~Schmidt, Phys. Rev. D \textbf{49}, 4510 (1994).

\bibitem{Stirling:1994he}
W.J.~Stirling, Nucl. Phys. B \textbf{423}, 56 (1994).

\bibitem{Vera:2006un}
  A.~Sabio Vera,
  Nucl.\ Phys.\ B {\bf 746} (2006) 1
  [hep-ph/0602250].

\bibitem{Vera:2007kn}
  A.~Sabio Vera and F.~Schwennsen,
  Nucl.\ Phys.\ B {\bf 776} (2007) 170
  [hep-ph/0702158 [HEP-PH]].

\bibitem{Khachatryan:2016udy}
  V.~Khachatryan {\it et al.} [CMS Collaboration],
  JHEP {\bf 1608} (2016) 139   [arXiv:1601.06713 [hep-ex]].

\bibitem{Caporale:2015vya}
  F.~Caporale, G.~Chachamis, B.~Murdaca, A.~Sabio Vera,
  Phys.\ Rev.\ Lett.\  {\bf 116} (2016) no.1,  012001
  [arXiv:1508.07711 [hep-ph]].

\bibitem{Caporale:2016soq}
  F.~Caporale, F.G.~Celiberto, G.~Chachamis, D.~Gordo Gomez, A.~Sabio Vera,
  Nucl.\ Phys.\ B {\bf 910} (2016) 374
  [arXiv:1603.07785 [hep-ph]].

\bibitem{Caporale:2016zkc}
  F.~Caporale, F.G.~Celiberto, G.~Chachamis, D.~Gordo Gomez, A.~Sabio Vera,
  Phys.\ Rev.\ D {\bf 95} (2017) no.7,  074007
    [arXiv:1612.05428 [hep-ph]].

\bibitem{Caporale:2015int}
  F.~Caporale, F.G.~Celiberto, G.~Chachamis, A.~Sabio Vera,
  Eur.\ Phys.\ J.\ C {\bf 76} (2016) no.3,  165
  [arXiv:1512.03364 [hep-ph]].

\bibitem{Caporale:2016xku}
  F.~Caporale, F.G.~Celiberto, G.~Chachamis, D.~Gordo Gomez, A.~Sabio Vera,
  Eur.\ Phys.\ J.\ C {\bf 77} (2017) no.1,  5
  [arXiv:1606.00574 [hep-ph]].
  
\bibitem{Celiberto:2016hae}
  F.G.~Celiberto, D.Yu.~Ivanov, B.~Murdaca, A.~Papa,
  Phys.\ Rev.\ D {\bf 94} (2016) no.3,  034013
  [arXiv:1604.08013 [hep-ph]].

\bibitem{hadrons}
  D.Yu.~Ivanov, A.~Papa,
  JHEP {\bf 1207} (2012) 045
  [arXiv:1205.6068 [hep-ph]].

\bibitem{Chirilli:2012jd}
   G.A.~Chirilli, B.W.~Xiao, F.~Yuan,
   Phys.\ Rev.\ D {\bf 86} (2012) 054005
   [arXiv:1203.6139 [hep-ph]].

\bibitem{Iancu:2016vyg}
   E.~Iancu, A.H.~Mueller, D.N.~Triantafyllopoulos,
   JHEP {\bf 1612} (2016) 041
   [arXiv:1608.05293 [hep-ph]].

\bibitem{Khachatryan:2010gv}
  V.~Khachatryan {\it et al.} [CMS Collaboration],
Proton-Proton Collisions at the LHC,''
  JHEP {\bf 1009} (2010) 091
  doi:10.1007/JHEP09(2010)091
  [arXiv:1009.4122 [hep-ex]].

\bibitem{Khachatryan:2015lva}
  V.~Khachatryan {\it et al.} [CMS Collaboration],
in pp collisions at $\sqrt s =$13 TeV,''
  Phys.\ Rev.\ Lett.\  {\bf 116} (2016) no.17,  172302
  doi:10.1103/PhysRevLett.116.172302
  [arXiv:1510.03068 [nucl-ex]].

\bibitem{Aad:2015gqa}
  G.~Aad {\it et al.} [ATLAS Collaboration],
$\sqrt{s}=$13 and 2.76 TeV $pp$ Collisions with the ATLAS Detector,''
  Phys.\ Rev.\ Lett.\  {\bf 116} (2016) no.17,  172301
  doi:10.1103/PhysRevLett.116.172301
  [arXiv:1509.04776 [hep-ex]].

\bibitem{Brodsky:1982gc}
S.J.~Brodsky, G.P.~Lepage, P.B.~Mackenzie, Phys. Rev. D \textbf{28}, 228 (1983).

\bibitem{Brodsky:1996sgBrodsky:1997sdBrodsky:1998knBrodsky:2002ka}
S.J.~Brodsky, F.~Hautmann, D.E.~Soper, Phys. Rev. Lett. {\bf 78}, 803
(1997). [Erratum: Phys. Rev. Lett. {\bf 79}, 3544 (1997)];
%
Phys. Rev. D {\bf 56}, 6957 (1997);
%
S.J.~Brodsky, V.S.~Fadin, V.T.~Kim, L.N.~Lipatov, G.B.~Pivovarov, 
JETP Lett. {\bf 70}, 155 (1999);
%
JETP Lett. {\bf 76}, 249 (2002).

\bibitem{Martin:2009iq}
  A.D.~Martin, W.J.~Stirling, R.S.~Thorne, G.~Watt,
  Eur.\ Phys.\ J.\ C {\bf 63} (2009) 189

\bibitem{Albino:2008fy}
S.~Albino, B.A.~Kniehl, G.~Kramer,
Nucl.\ Phys.\ B {\bf 803}, 42 (2008).

\bibitem{Hirai:2007cx}
M.~Hirai, S.~Kumano, T.-H.~Nagai, K.~Sudoh,
Phys.\ Rev.\ D {\bf 75}, 094009 (2007). 

\bibitem{Kotikov:2000pm} 
A.V.~Kotikov and L.N.~Lipatov,
Nucl.\ Phys.\ B {\bf 582}, 19 (2000).

\bibitem{Harland-Lang:2014zoa}
  L.A.~Harland-Lang, A.D.~Martin, P.~Motylinski, R.S.~Thorne,
  Eur.\ Phys.\ J.\ C {\bf 75} (2015) no.5,  204
  [arXiv:1412.3989 [hep-ph]].

\bibitem{Dulat:2015mca}
  S.~Dulat et al. 
  Phys.\ Rev.\ D {\bf 93} (2016) no.3,  033006
  [arXiv:1506.07443 [hep-ph]].

\bibitem{DSS}
D.~de Florian, R.~Sassot, M.~Stratmann,
Phys. Rev. D {\bf 75}, 114010 (2007); Phys. Rev. D {\bf 76}, 074033 (2007).

\bibitem{cernlib}
CERNLIB Homepage: \url{http://cernlib.web.cern.ch/cernlib}.

\bibitem{chyp}
R.~Forrey, J. Comput. Phys. {\bf 137} (1997) 79.

\bibitem{rpsi}
W.J.~Cody, A.J.~Strecok, H.C.~Thacher, Math. Comput. {\bf 27} (1973) 121.

\bibitem{Yang:2015avi} 
  D.J.~Yang, F.J.~Jiang, W.C.~Chang, C.W.~Kao, S.~i.~Nam,
  Phys.\ Lett.\ B {\bf 755}, 393 (2016).

\bibitem{Owens:2001rr} 
  J.F.~Owens,
  Phys.\ Rev.\ D {\bf 65}, 034011 (2002).

\bibitem{Almeida:2009jt} 
  L.G.~Almeida, G.F.~Sterman, W.~Vogelsang,
  Phys.\ Rev.\ D {\bf 80}, 074016 (2009).

\bibitem{Diehl:2011yj}
M.~Diehl, D.~Ostermeier, A.~Schafer,
JHEP {\bf 1203}, 089 (2012).
  
\bibitem{Maciula:2014pla} 
R.~Maciula, A.~Szczurek,
Phys.\ Rev.\ D {\bf 90}, no. 1, 014022 (2014).

\end{thebibliography}
\end{document}